\documentclass[reprint,amsmath,amssymb,aps,prb]{revtex4-1}

\usepackage{graphicx}
\usepackage{dcolumn}
\usepackage{scrextend}
\usepackage[colorlinks=true, citecolor=blue, linkcolor=red]{hyperref}
\usepackage{bm}
\setcitestyle{numbers,square}
\usepackage{longtable}

\begin{document}

\preprint{APS/123-QED}

\title{Intrinsic Bending Flexoelectric Constants in Two-Dimensional Materials}

\author{Xiaoying Zhuang$^{*,\dagger,\P}$, Bo He$^{\dagger}$, Brahmanandam Javvaji$^{\dagger}$ and Harold S. Park$^{*,\ddag}$ \\
	\emph{\small{$^{\dagger}$Institute of Continuum Mechanics, Leibniz Universit\"at Hannover, Appelstr. 11, 30167 Hannover, Germany}} \\
	\emph{\small{$^{\ddag}$ Department of Mechanical Engineering, Boston University, Boston, Massachusetts 02215, USA}} \\
	\emph{\small{$^{\P}$College of Civil Engineering, Tongji University, 1239 Siping Road, 200092 Shanghai, China}} \\
	\small{$^*$E-mail:~zhuang@ikm.uni-hannover.de; parkhs@bu.edu} \\
	\small{Phone:+49 511 762 19589. Fax: +49 511 762 5496}
}
%
%
%
%
\date{\today}

\begin{abstract}
Flexoelectricity is a form of electromechanical coupling that has recently emerged because, unlike piezoelectricity, it is theoretically possible in any dielectric material.  Two-dimensional (2D) materials have also garnered significant interest because of their unusual electromechanical properties and high flexibility, but the intrinsic flexoelectric properties of these materials remain unresolved.  In this work, using atomistic modeling accounting for charge-dipole interactions, we report the intrinsic flexoelectric constants for a range of two-dimensional materials, including graphene allotropes, nitrides, graphene analogs of group-IV elements, and the transition metal dichalcogenides (TMDCs).  We accomplish this through a proposed mechanical bending scheme that eliminates the piezoelectric contribution to the total polarization, which enables us to directly measure the flexoelectric constants. While flat 2D materials like graphene have low flexoelectric constants due to weak $\pi-\sigma$ interactions, buckling is found to increase the flexoelectric constants in monolayer group-IV elements.  Finally, due to significantly enhanced charge transfer coupled with structural asymmetry due to bending, the TMDCs are found to have the largest flexoelectric constants, including MoS$_{2}$ having a flexoelectric constant ten times larger than graphene. 
\end{abstract}

\pacs{Valid PACS appear here}
\maketitle


\section{Introduction}
\label{sec:intro}
\noindent  Piezoelectricity is perhaps the best-known mechanism of converting mechanical deformation into electrical energy, and has been widely used in engineering practice \cite{ikeda1996fundamentals,heywang2008piezoelectricity}.  While piezoelectricity is well-established, other types of electromechanical coupling, such as flexoelectricity, have recently attracted significant interest \cite{ma2006flexoelectricity,harden2006giant,hong2010flexoelectricity,kalinin2008electronic,krichen2016flexoelectricity}.  One reason for this is that piezoelectricity is limited to materials with non-centrosymmetric crystal structures.  In contrast, in flexoelectricity the polarization is not only related to the strain as in piezoelectricity, but also the strain gradient.  Thus, for flexoelectricity, the polarization $P$ induced due to mechanical deformation is given as \cite{Tagantsev1986}
\noindent 
\begin{equation}
P^{\alpha} = d^{\alpha \beta \gamma}\epsilon^{\beta \gamma} + \mu^{\alpha \beta \gamma \delta}\frac{\partial\epsilon^{\gamma \delta}}{\partial x_{\beta}},
\label{eq:polarization}
\end{equation}
where $d^{\alpha \beta \gamma}$ is the piezoelectric coefficient, $\epsilon^{\beta \gamma}$ is the strain, $\mu^{\alpha \beta \gamma \delta}$ is the flexoelectric coefficient, $\frac{\partial\epsilon^{\gamma \delta}}{\partial x_{\beta}}$ is the strain gradient and $\alpha$, $\beta$, $\gamma$ and $\delta$ represent the directional components of the coordinate system. 

Because flexoelectricity is dependent on the gradient of strain it can, in principle, occur in any dielectric material.  Furthermore, significant potential for flexoelectricity emerges as the dimensions of materials reduce to the nanometer scale due to the ability to produce larger strain gradients for small size scales.  However, studies of electromechanical coupling in nanomaterials, and specifically two-dimensional (2D) materials such as graphene and molybdenum disulfide (MoS$_{2}$) have largely focused on their piezoelectric properties~\cite{Wang2015observattion,Song2017piezo,Brennan2017out,Zheng2017hexagonal,Zelisko2014anomalous,blonsky2015ab,Kundalwal2017,chandratre2012coaxing,duerloo2013flexural,zhang2017boron,zhou2016theoretical,javvaji2018generation,wu2014piezoelectricity,duerloo2012intrinsic}; we also note a recent review article summarizing the various simulation and experimental methods for characterizing piezoelectricity in 2D materials~\cite{hinchet2018piezoelectric}.  

In contrast to the extensive study of piezoelectricity in nanomaterials, relatively few studies on flexoelectricity have been performed.  Majdoub and co-workers reported an enhancement of flexoelectricity in nanoscale Barium Titanium oxide (BTO)~\cite{majdoub2008enhanced}.  Surface effects on flexoelectricity in BTO nanobelts were investigated using core-shell potentials \cite{He2018}. In addition, several preliminary studies on flexoelectricity in 2D materials have recently been carried out using density functional theory (DFT) calculations, theoretical analyses or experiments.  For instance, a linear relationship was found between induced dipole moment and bending curvature in graphene using DFT calculations~\cite{kalinin2008electronic}.  A theoretical analysis~\cite{kvashnin2015flexoelectricity} of flexoelectricity in carbon nanostructures (nanotubes, fullerenes and nanocones) confirmed the dependence of flexoelectric atomic dipole moments on local curvature.  Others have patterned graphene to generate strain gradients and enhance the electromechanical coupling and polarization~\cite{chandratre2012coaxing,Kundalwal2017,javvaji2018generation}. Furthermore, a recent experimental study~\cite{Brennan2017out} provided evidence that monolayer MoS$_2$ exhibits an out-of-plane flexoelectric response using the piezoresponse force microscopy. However, one key issue in calculating or measuring the flexoelectric constants of 2D materials is that it has been difficult to isolate the relative contributions of piezoelectricity and flexoelectricity to the resulting polarization~\cite{Brennan2017out}.  As a result, the intrinsic flexoelectric properties of 2D materials remain unresolved, and furthermore the mechanisms controlling the intrinsic flexoelectric properties of different 2D materials are also unresolved.  

\begin{figure}
	\centering
	\includegraphics[width=\linewidth]{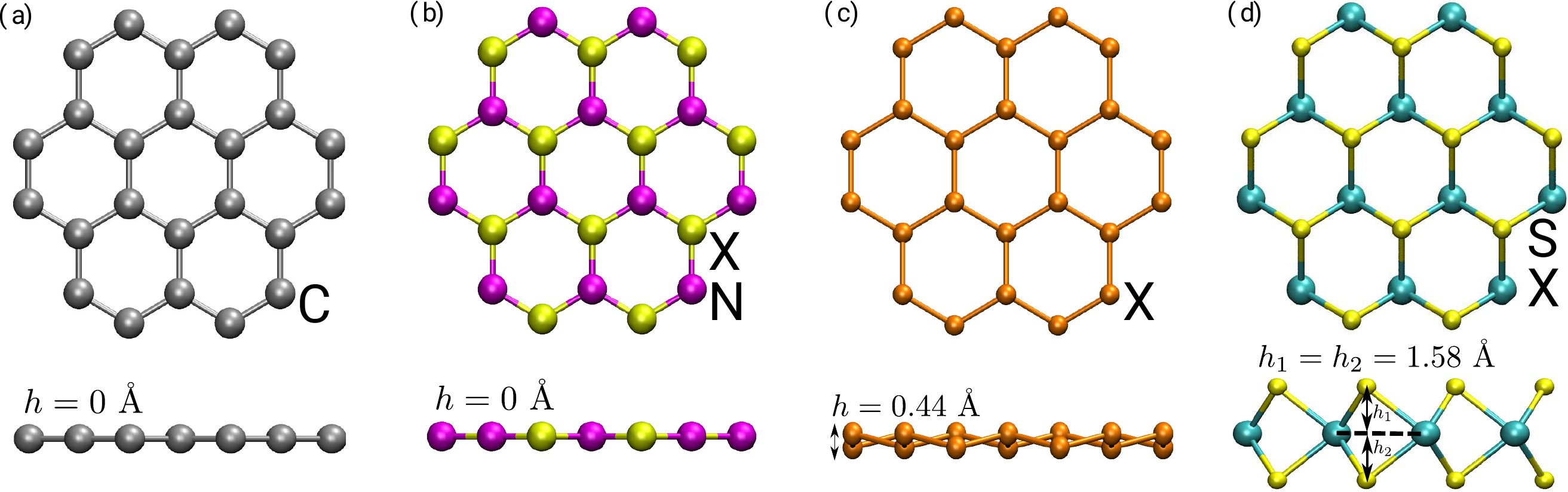}
	\caption{Top and side view of the studied materials: (a) Graphene allotropes; (b) Nitrides XN, X = B, Al, Ga; (c) graphene analogues of group-IV elements X, X = B, Si, Ge; (d) transition metal dichalcogenides XS$_2$, X = Cr, Mo, W.  For (a)-(c), $h$ refers to the buckling height, while in (d) $h_{1}$ and $h_{2}$ refer to intralayer distances.}  
	\label{Fig:2d_material_geometry}
\end{figure}

In this work, we develop a classical charge-dipole (CD) atomistic model that couples with classical molecular dynamics (MD) simulations to calculate the intrinsic bending flexoelectric constants of the four different 2D material groups shown in Fig. \ref{Fig:2d_material_geometry}:  graphene allotropes, nitrides, graphene analogues of group-IV elements and transition metal dichalcogenides (TMDCs).  Specifically, we propose and validate a mechanical bending formulation that eliminates the piezoelectric contribution to the polarization in Eq. (\ref{eq:polarization}), thus enabling us to directly calculate the intrinsic flexoelectric constants. By comparing these different classes of 2D materials, we investigate and elucidate the effects of charge-dipole interactions, out of plane buckling in monolayers, and intralayer buckling asymmetry and charge transfer on the flexoelectric response of 2D materials.

\section{Simulation method}
\label{sec:simulation_method}
\noindent In this work, we performed MD simulations by utilizing a CD model in conjunction with bonded interactions to determine the atomic configurations, as well as the point charges $q_{i}$ and dipole moments $\mathbf{p}_{i}$ associated with each atom $i$.  The bonded interactions were modeled using well-known potentials, i.e. adaptive intermolecular reactive empirical bond order (AIREBO), Tersoff and Stillinger Weber (\textcolor{blue}{see Table \ref{table:charge_width_detail} for references to all potentials}), while the point charges and dipole moments were calculated using the well-known CD potentials \cite{olson1978atom,Mayer2007,Mayer2005} (further details are given in \textcolor{blue}{Appendix}  \ref{sec:Charge-diople potential model}). The potential parameters for the CD model were determined using DFT calculations (more details are given in \textcolor{blue}{Appendix} \ref{sec:Est-R}), and were validated through calculation of piezoelectric constants for boron nitride and MoS$_{2}$, which as shown in Table \ref{table:piezo_validate} in \textcolor{blue}{Appendix} \ref{sec:Validation_CD} are in good agreement with previous studies. All simulations were performed using the open-source MD simulation code large-scale atomic/molecular massively parallel simulator (LAMMPS) \cite{Plimpton1995}.

\begin{center}
	\begin{figure}[h]
		\begin{center}
			\includegraphics[scale=0.5]{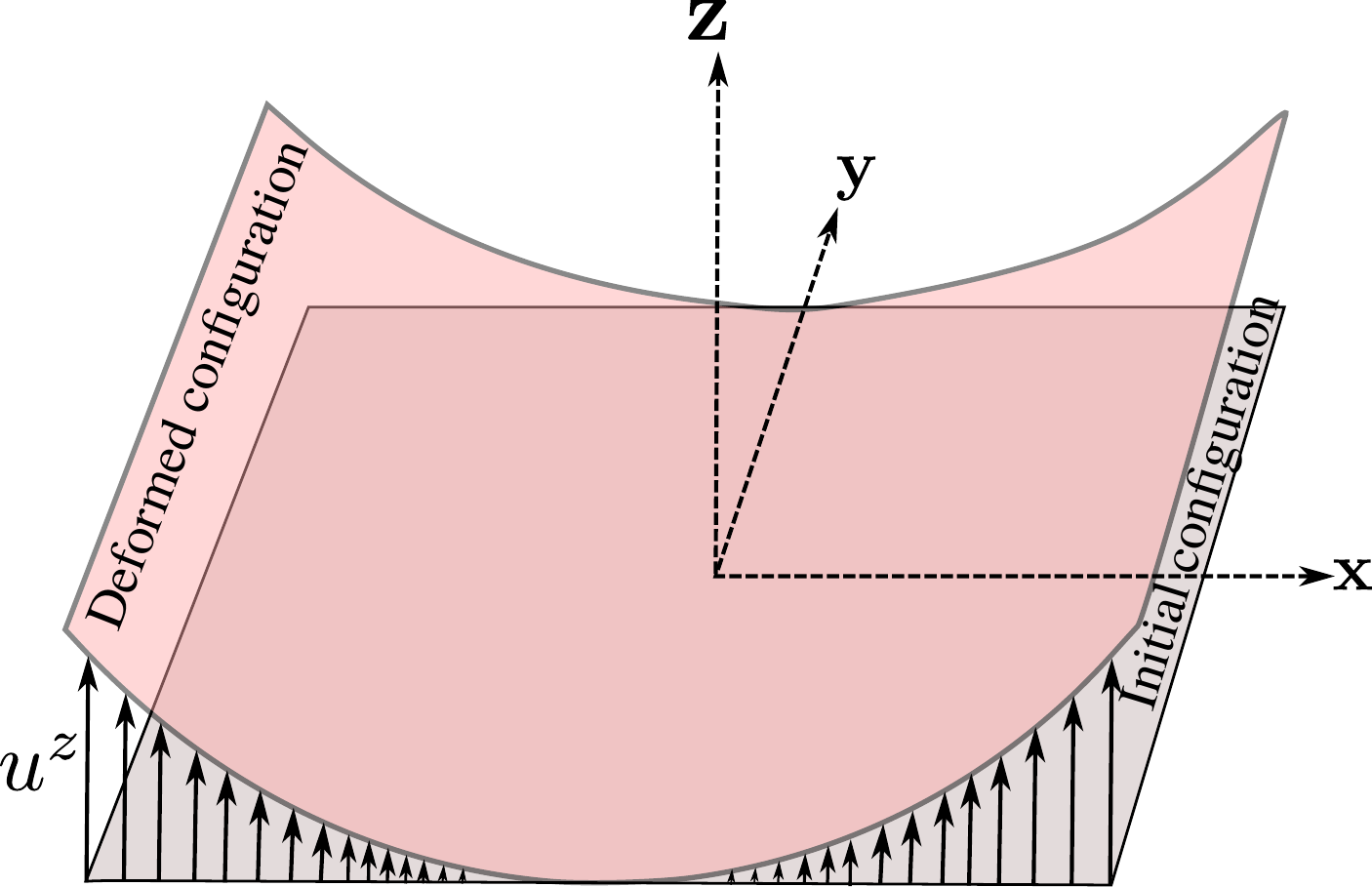}
		\end{center}
		\caption{Schematic illustration of geometry and loading condition for 2D material system.}
		\label{fig:bending-model}
	\end{figure}
\end{center}

The MD simulations were performed using the unit cell dimensions ($a, b, c$ and $h$) for each material given in  Table \ref{table:charge_width_detail}, along with the potential functions employed to estimate the bonded interactions.  A fixed unit cell size of $80\times 80$~\AA \: \cite{sizeinfo} was adopted for all simulations to estimate the flexoelectric coefficients.  The flexoelectric constants were determined by first prescribing the following displacement field to the atomic system
\begin{equation}
u^z = K\frac{x^2}{2}, 
\label{eq:out-bending}
\end{equation}
where $x$  represents the atom coordinate in the $x$ direction, $K$ represents the inverse of curvature (strain gradient) of the bending plane, and where the prescribed mechanical deformation is shown in 
Fig.\ref{fig:bending-model}.  Once the bending deformation is prescribed, the edge region atoms are held fixed while the interior atoms are allowed to relax to energy minimizing positions using the conjugate-gradient algorithm, after which the point charges and dipole moments are found for each atom.

From the MD simulations, we establish the relationship between polarization and strain gradient as follows. The strain gradient from Eq.~(\ref{eq:out-bending}) is
\begin{equation}
\frac{\partial \varepsilon^{xz}}{\partial x} = \textcolor{blue}{\frac{1}{2}}\frac{\partial^2 u^{z}}{\partial x^2} = \textcolor{blue}{\frac{1}{2}} K
\label{eq:out-sgrad}
\end{equation}
where $\varepsilon^{xz}$ is the strain in the $x$ direction from the applied deformation in the $z$ direction. Substituting Eq.~(\ref{eq:out-sgrad}) in Eq.~(\ref{eq:polarization}) and assuming that the imposed mechanical deformation in Eq.~(\ref{eq:out-bending}) removes the piezoelectric contribution, we obtain
\begin{equation}
P^{z} = \textcolor{blue}{\frac{1}{2}} \mu^{zxzx} K
\label{eq:out-flexo}
\end{equation}
where $\mu^{zxzx}$ is the out-of-plane or bending flexoelectric coefficient and $P^{z}$ is the out-of-plane polarization. We will verify the assumption of the removal of the piezoelectric contribution through the prescribed bending deformation in the next section.

\section{Results and discussion}
\label{sec:result}
We use the simulation procedure described previously in Sec. \ref{sec:simulation_method} to study the flexoelectric properties of four groups of 2D materials: graphene allotropes (C1, C2 and C3), nitrides (BN, AlN and GaN), graphene analogues of group-IV elements (Si, Ge and Sn) and TMDC monolayers (MoS$_2$, WS$_2$ and CrS$_2$). C1 corresponds to pristine graphene, while C2 and C3 represent graphene with Stone-Wales defects which replace some hexagons by pentagons and heptagons with different periodicity, respectively \cite{enyashin2011graphene}. BN, AlN and GaN are the nitrogen-based hexagonal monolayers with boron, aluminum and gallium, respectively. Silicene (Si), Germanene (Ge) and Stanene (Sn) are the group-IV 2D graphene analogs. However, the vertical distance between the atoms or buckling height $(h)$ in the unit cell is non-zero when compared to the graphene allotropes and nitride material groups (see Fig.~\ref{Fig:2d_material_geometry}(a) and (c)).  The TMDCs possess three sub layers or intra-layers where element 'X' (center layer) forms bonds with two S atoms in the top and bottom layers.  The layers are vertically separated by the intralayer heights $h_1$ and $h_2$, as shown in Fig.~\ref{Fig:2d_material_geometry}(d).  

We first demonstrate that the proposed bending scheme eliminates the piezoelectric contribution to the total polarization, such that we can focus on the resulting intrinsic flexoelectric properties of the different 2D material groups.  The applied deformation (using Eq.~(\ref{eq:out-bending})) results in strain $(\epsilon^{xz})$ and strain gradient $(\frac{\partial \epsilon^{xz}}{\partial x})$ along the $xz$ direction, and a polarization along the $z$ direction, where we use MoS$_{2}$ as an example as it has the most complex 2D structure of the 2D materials we consider.  We calculate the local atomic strain for each atom $i$ using the local deformation gradient $\mathbf{F}$ which involves the initial and deformed atomic coordinates. The local atomic strain tensor for atom $i$ $(\epsilon_i)$ is \cite{Javvaji2017stable}
\begin{equation}
\epsilon_i = \frac{1}{2} \left[ \left(\mathbf{F}_i\right)^{\text{T}} \mathbf{F}_i - \mathbf{I}\right],
\label{eq:astrain}
\end{equation}
where $\mathbf{I}$ is the identity matrix. 

\begin{figure}
	\centering
	\includegraphics[width=1.0\linewidth]{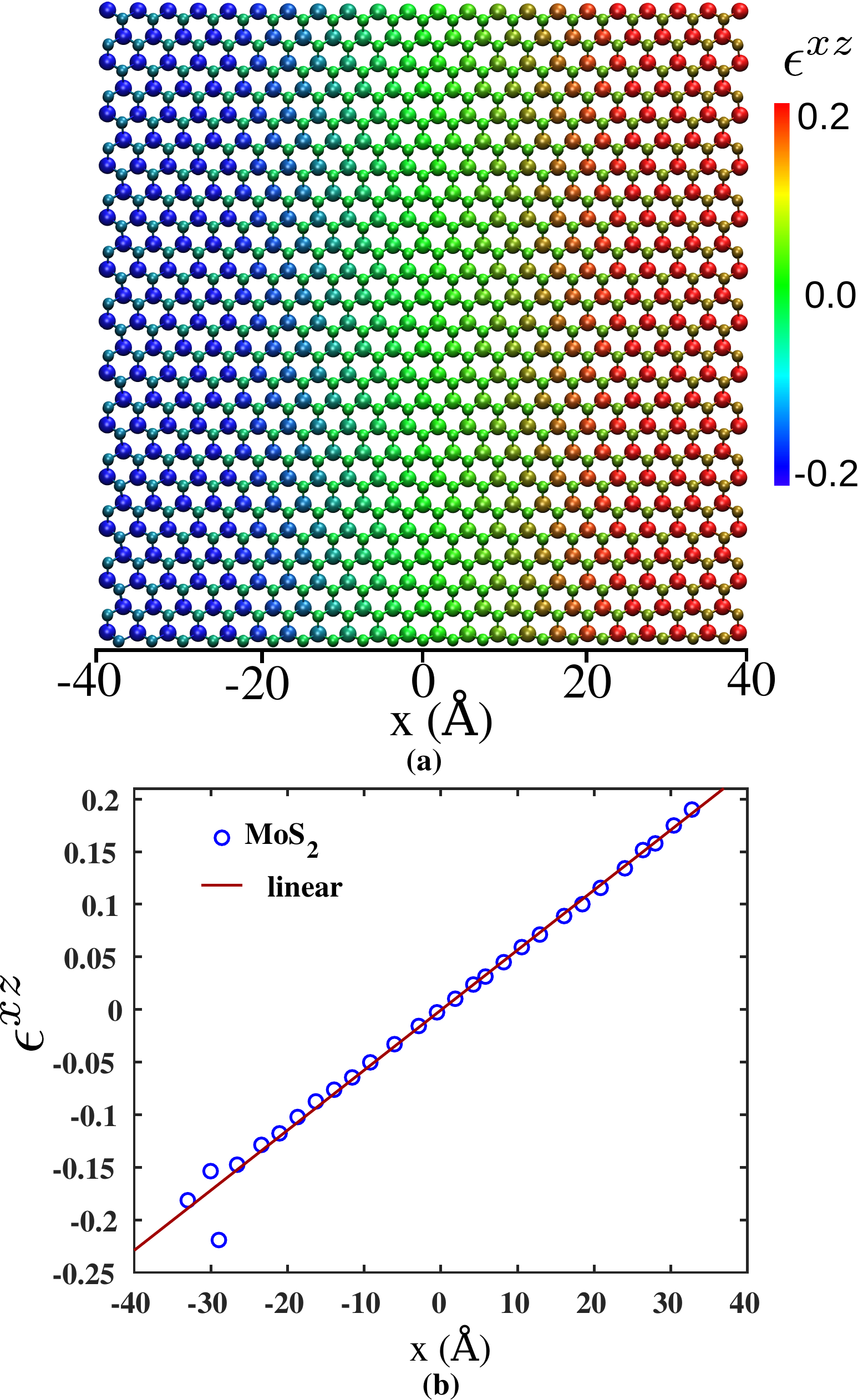}
	\caption{(a) Atomic configuration colored with strain $\epsilon^{xz}$ in $x$ direction for MoS$_2$ sheet when strain gradient $K$ = 0.01 $\text{\AA}^{-1}$; the large spheres represent Mo atoms and small spheres represent S atoms.   (b) Bin-wise distribution of strain $\epsilon^{xz}$ along $x$ axis, circles represent the calculated average strain $\epsilon^{xz}$ at location $x$ and solid line is linear fitting to the calculated data. }
	\label{fig:astrain-epsxz}
\end{figure}

Fig.~\ref{fig:astrain-epsxz}(a) represents the atomic configuration of MoS$_2$ system colored with the $xz$ component of strain, which is calculated from Eq.~(\ref{eq:astrain}) at a given curvature $(K=0.01 ~\text{\AA}^{-1})$. The variation of strain $\epsilon^{xz}$ along the $x-$ direction is plotted in Fig.~\ref{fig:astrain-epsxz}(b), where the strain was found by dividing the atomic system into several equal width bins and averaging the strain in each bin.  
A linear variation in $\epsilon^{xz}$ is observed from Fig.~\ref{fig:astrain-epsxz}(b).  This demonstrates that the induced deformation is symmetric and the resulting polarization due to strain is canceled out.  Therefore, the total strain $\epsilon^{xz}$ is zero (sum over all the bins), which eliminates the piezoelectric contribution to the polarization in Eq.~(\ref{eq:polarization}) and supports the assumption made in obtaining Eq.~(\ref{eq:out-flexo}), i.e. that for the prescribed bending deformation, the \textcolor{blue}{out-of-plane} polarization is only dependent on the strain gradient. \textcolor{blue}{Furthermore, symmetry analysis on the piezoelectric tensor show that $d^{zxz}$ is zero for a point group symmetry associated with the 2D material sets \cite{DeJong2015}.}

\textcolor{blue}{The mechanical bending deformation that is imposed serves to strictly to zero out the out-of-plane piezoelectric contribution to total polarization.  However, it is important to note that an in-plane polarization is generated due to the out-of-plane bending.  Furthermore, the in-plane polarization from out-of-plane bending may receive a contribution from in-plane piezoelectricity.  We further discuss the in-plane polarization in Section~\ref{sec:tmds}  and Appendix~\ref{sec:in-plane}.}

\subsection{Mechanisms of inducing polarization in 2D materials}

Our analysis of the mechanisms governing the flexoelectric constants for the 2D materials depends on understanding, within the framework of the utilized CD model, the various contributions to the dipole moments that are induced from the prescribed bending deformation.  Specifically, the dipole moment $\mathbf{p}_{i}$ on atom $i$ depends on its polarizability and the presence of a local electric field, which consists of three parts:  the electric field at position $\mathbf{r}_i$ due to neighboring (i) dipoles $\mathbf{p}_j$; (ii) charges $q_j$ and (iii) from the externally applied electric fields $\mathbf{E}^{\text{ext}}$.  Because the only external stimulus is the prescribed bending deformation, $\mathbf{E}^{\text{ext}}=0$ and the governing equation for the dipole moments (Eq.~(\ref{eq:GDEp2})) becomes
\begin{equation}
\mathbf{T}_{ii}^{p-p} \mathbf{p}_i - \sum_{j,i\neq j}^{N} \mathbf{T}_{ij}^{p-p} \mathbf{p}_j = \sum_{j,i\neq j}^{N} \mathbf{T}_{ij}^{q-p} q_j.
\label{eq:dpmi_2}
\end{equation}
where $\mathbf{T}_{ij}^{p-p}$ and $\mathbf{T}_{ij}^{q-p}$ are the polarizability tensors.  These two tensors represent dipole-dipole and charge-dipole interactions, respectively, which can also be interpreted as accounting for $\sigma-\sigma$ and $\sigma-\pi$ electron interactions, respectively\cite{Mayer2007,Mayer2005,Robert2014}, and can be written as\cite{Mayer2005},
\begin{widetext}
	\begin{align}
		\mathbf{T}_{ij}^{q-p} &= \frac{1}{4\pi\epsilon_0} \frac{\mathbf{r}_{ij}}{r_{ij}^3} \approx \frac{1}{4\pi\epsilon_0} \frac{\mathbf{r}_{ij}}{r_{ij}^3} \left[ \text{erf}\left(\frac{r_{ij}}{\sqrt{2}R}\right)-\sqrt{\frac{2}{\pi}} \frac{r_{ij}}{R} \text{exp}\left(-\frac{r_{ij}^2}{2R^2}\right)\right], \label{eq:Tqpij} \\
		\mathbf{T}_{ij}^{p-p} &= \frac{1}{4\pi\epsilon_0} \frac{3 \mathbf{r}_{ij} \otimes \mathbf{r}_{ij} - r_{ij}^2 \mathbf{I}}{r_{ij}^5} \left[ \text{erf}\left(\frac{r_{ij}}{\sqrt{2}R}\right)-\sqrt{\frac{2}{\pi}} \frac{r_{ij}}{R} \text{exp}\left(-\frac{r_{ij}^2}{2R^2}\right)\right] - \frac{1}{4\pi\epsilon_0} \sqrt{\frac{2}{\pi}} \frac{\mathbf{r}_{ij} \otimes \mathbf{r}_{ij}}{r_{ij}^2} \frac{1}{R^3} \text{exp}\left(-\frac{r_{ij}^2}{2R^2}\right).
		\label{eq:Tppij}
	\end{align}
\end{widetext}
From Eqs.~(\ref{eq:Tqpij}) and (\ref{eq:Tppij}), the inter-atomic distance $(r_{ij})$ and $R$ (\textcolor{blue}{factor related to polarizability}) are identified as the important factors in defining the dipole moment of atoms via the polarizability tensors.

\subsection{Flat 2D Monolayers}

We first consider the simplest 2D structures, flat graphene and BN monolayers.  To aid in the analysis, we 
rewrite Eq.~(\ref{eq:dpmi_2}) for only the $p_{i}^{z}$ component, which is
\begin{equation}
T_{ii}^{p-p,zz} p_i^z = E_{i}^{p,z} + E_{i}^{q,z}
\label{eq:dmpi_3}
\end{equation}
where $E_i^{p,z}=\sum_{j,i\neq j}^N \lbrace T_{ij}^{p-p,xz} p_j^x + T_{ij}^{p-p,yz} p_j^y + T_{ij}^{p-p,zz} p_j^z \rbrace $ and $E_i^{q,z}= \sum_{j,i\neq j}^N T_{ij}^{q-p,z} q_j$ are the electric fields on atom $i$ due to neighboring dipoles, charges and associated polarizability components. 

\begin{figure}
	\centering
	\includegraphics[scale=0.5]{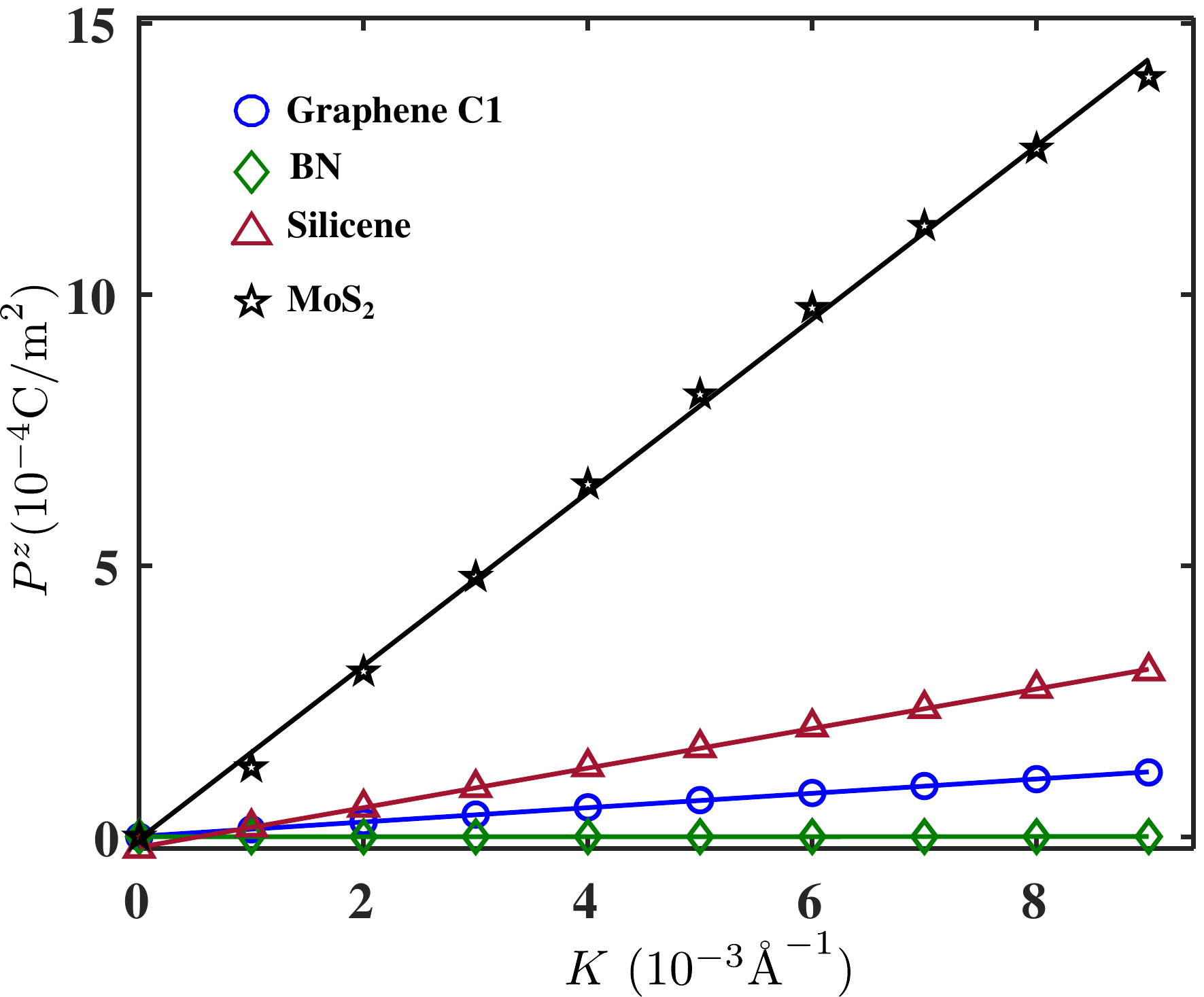}
	\caption{ Polarization $P^z$ vs strain gradient $K$ for graphene, BN, Silicene and MoS$_2$. Markers indicate the simulation data and solid lines indicate the linear fitting.} 
	\label{fig:Fx_Fz_x}
\end{figure}

For the undeformed graphene sheet, the out of plane dipole moment $p_{i}^{z}$ is zero due to the flat nature of the monolayer.  However, once graphene is bent, the $\pi-\sigma$ interactions increase, leading to a non-zero $p_{i}^{z}$.  Specifically, for deformed graphene with bending curvature $0.002~\text{\AA}^{-1}$, the measured contributions of $E_i^{p,z}$ and $E_i^{q,z}$ to the total electric field on atom $i$ are $93.45$ and $6.55\%$, respectively.  As graphene is bent further, these contributions change to $93.27$ and $6.73\%$, respectively, when $K = 0.01~\text{\AA}^{-1}$.  The increased importance of $E_{i}^{q,z}$ with increasing bending implies an increasing importance of $\pi-\sigma$ interactions on the total electric field and dipole moment on atom $i$.  This can also be interpreted through pyramidalization~\cite{Surya2012,dumitricua2002curvature,Nikiforov2014}, in which sp$^2$ bonding converts to sp$^3$ bonding.  In this process, the valence electrons of each carbon atom develops bonding interactions with neighboring atoms due to the bond bending involved symmetry reduction, which allows mixing between $\pi$ and $\sigma$ electrons, leading to $\pi-\sigma$ interactions~\cite{Gleiter1987}.  This interaction modifies the charge state of the carbon atom as well as the locally generated electric fields, which is captured by the CD model in the form of the charge-induced electric fields $E^{q,z}$.  Overall, these increased $\pi-\sigma$ interactions result in the flexoelectric coefficient for graphene being found as $\mu_{\text{gr}}=0.00286$ nC/m, which is found from linear fitting of the polarization as a function of bending curvature in Fig.~\ref{fig:Fx_Fz_x}.

\begin{figure}[!htbp]
	\centering	
	\includegraphics[width=1.0\linewidth]{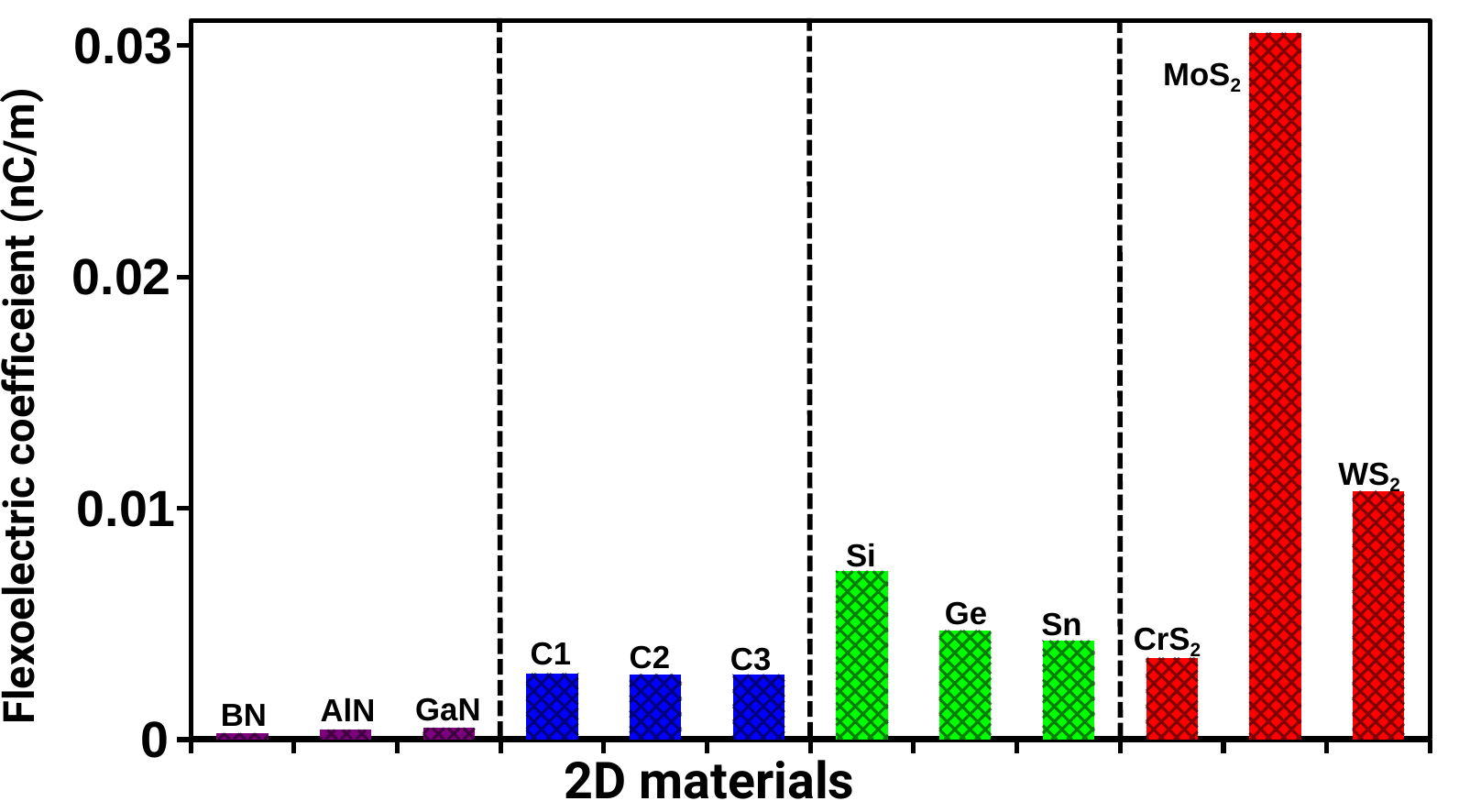}
	\caption{Bending flexoelectric coefficient for 2D materials}
	\label{fig:flexo_coef_summary}
\end{figure}

In case of BN, the contribution of $E^{q,z}$ also increases from $1.86$ to $1.99\%$ when bending curvature increases from $0.002$ to $0.01~\text{\AA}^{-1}$, though the overall contribution of $E^{q,z}$ to the total electric field is smaller than for graphene.  This suggests that the $\pi-\sigma$ interactions in BN are weaker than in graphene, which may be related to the difference in the tendency of pyramidalization between B and N atoms.  Specifically, B atoms prefer the sp$^2$ hybridization while N atoms are more likely to achieve sp$^3$ hybridization or pyramidalization \cite{Hernandez1998,Hernandez1999}.  Thus, even though the polarizability of BN is similar to graphene (see $\alpha_{\text{total}}^{CAL}$ values in Table.~\ref{table:charge_width_detail}), the flexoelectric constant of BN of 0.00026 nC/m is ten times smaller than graphene due to the smaller \textcolor{blue}{$E^{q,z}$} contribution in BN. 

The graphene allotropes C2 and C3 show similar flexoelectric coefficients to defect-free monolayer graphene (C1), as shown in Fig.~\ref{fig:flexo_coef_summary}.  Though C2 and C3 contain different arrangements of defects, the sp$^2$ hybridization is unchanged, which induces nearly equal charges and dipole moments for atoms in C2 and C3 under deformation. As a result, the flexoelectric coefficients are nearly constant for this material group. In the case of the nitride group, AlN and GaN are found to have larger flexoelectric constants than BN as shown in Fig.~\ref{fig:flexo_coef_summary}, though still significantly smaller than graphene.  This is due to a corresponding increase in the contribution of \textcolor{blue}{$E^{q,z}$}, from 1.99\% for BN to 2.25\% for AlN to 6.85\% for GaN for a curvature of $0.01~\text{\AA}^{-1}$. 

\subsection{Buckled 2D Monolayers} 

As seen in Fig.~\ref{fig:Fx_Fz_x}, the induced polarization for flat 2D materials is much smaller than is seen in silicene.  From a structural point of view, silicene and graphene differ in that the atomic polarizability of silicene is larger, and also that it exists in a buckled configuration as compared to graphene (see $h$ values in Table~\ref{table:charge_width_detail}).  Therefore, we performed simulations to examine the effects of both of these factors on the induced polarization in silicene.  We first performed a bending test for silicene in which the buckling height was kept to zero, in order to understand the effect of buckling on the polarization.  To do so, we simply imposed the bending deformation on silicene without allowing any subsequent relaxation of the atomic positions.  The variation of polarization for flat silicene and silicene is plotted in Fig.~\ref{fig:si-suppl}. From the numerical fitting, the flexoelectric coefficients for flat silicene $\mu_{\text{si-flat}}$ and silicene $\mu_{\text{si}}$ are identified as $0.00634$ and $0.00728$ nC/m, respectively.  Noting that the graphene flexoelectric coefficient is $\mu_{\text{gr}}=0.00286$ nC/m, the ratio of $\mu_{\text{si-flat}}/\mu_{\text{gr}}$ is $2.217$, which is close to the ratio of their atomic polarizability parameters ($R_{\text{si}}/R_{\text{gr}}=2.141$ from  Table~\ref{table:charge_width_detail}). From this, it is clear that the atomic polarizability increases the induced polarization and thus flexoelectric constants. The ratio of $\mu_{\text{si}}/\mu_{\text{gr}}$ is $2.545$, which is about 15\% higher than $2.217$. This increase in polarization of about 15\% between silicene and flat silicene can therefore be ascribed to the buckled structure of silicene.

Further understanding can be drawn from the contributions of the electric fields from dipole-dipole and charge-dipole interactions. The contributions from $E_i^{p,z}$ and $E_i^{q,z}$ to the total electric field are estimated for flat silicene and silicene when the bending curvature is $0.008~\text{\AA}^{-1}$. The numerical values for flat silicene are $91.89$ and $8.10\%$, respectively, which are similar to that of graphene. Therefore, the increased dipole moment and flexoelectric coefficient for flat silicene is primarily due to its larger atomic polarizability ($R_{\text{si}}/R_{\text{gr}}=2.141$) compared to graphene.

For buckled silicene, the contributions from $E_i^{p,z}$ and $E_i^{q,z}$ to the total electric field are $76.89$ and $23.10\%$, respectively. Comparing to flat silicene, there is an increase in $E_i^{q,z}$ and decrease in $E_i^{p,z}$ for silicene.  Thus, the CD model predicts that $\pi-\sigma$ interactions are dramatically enhanced in buckled silicene as compared to flat silicene, which is in agreement with recent DFT studies by Podsiad\l{}y-Paszkowska \textit{et. al.} \cite{Podsiady-Paszkowska2017}, who found that it is easier to achieve sp$^3$ bonding in buckled silicene.  Such changes in hybridization (pyramidalization) lead to significant charge modulations and induce large dipole moments. The difference in the numerical contribution of $E^{q,z}$ to the total electric field in flat vs. buckled silicene of 15\% is identical to the observed difference in magnitude of the flexoelectric coefficients.  This demonstrates that buckling in the atomic structure of 2D materials can induce increased polarization, and thus flexoelectric constants.  

Germanene and stanene also have flexoelectric constants that are larger than  graphene and BN as shown in Fig.~\ref{fig:flexo_coef_summary}, though lower than silicene.  This is due to a combination of lower polarizability of these materials as compared to silicene (see $\alpha_{\text{total}}^{CAL}$ values in Table.~\ref{table:charge_width_detail}), and due to reduced $E_{q,z}$ contributions of $20.91$ and $18.45\%$, respectively, indicating weaker $\pi-\sigma$ interactions in these buckled structures as compared to silicene.

\subsection{TMDCs}
\label{sec:tmds}
As shown in Fig.~\ref{fig:Fx_Fz_x}, the polarization under bending is significantly higher in MoS$_{2}$ than the other 2D materials.  Interestingly, for MoS$_{2}$, the contributions from the dipole and charge-induced electric fields in the $z-$direction are $15.23$ and $84.76\%$, respectively, where the contribution of $E^{q,z}$ is significantly higher than for the previously discussed 2D materials. 

\begin{figure}
	\centering
	\includegraphics[scale=0.4]{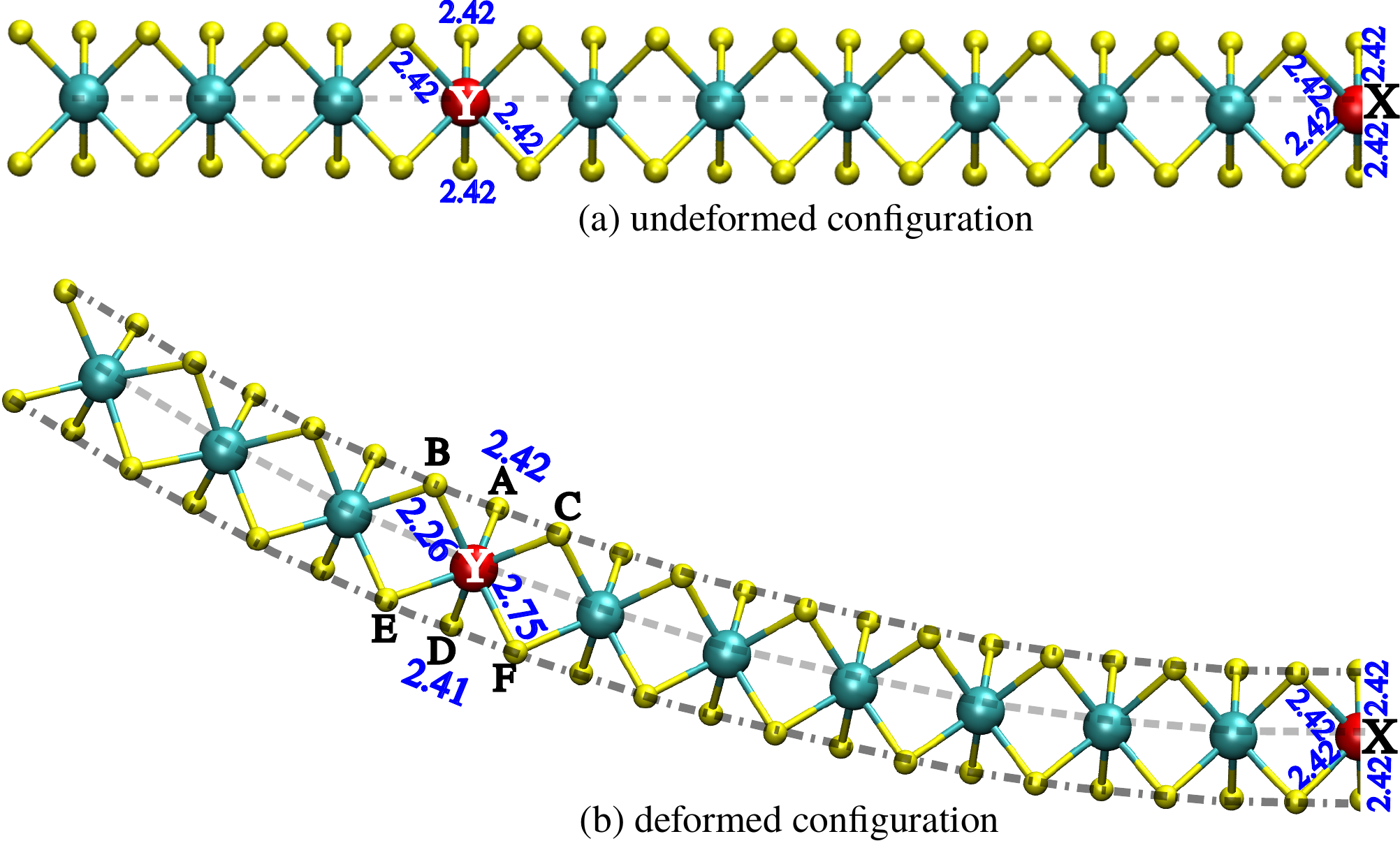}
	\caption{(a) Undeformed and (b) deformed atomic configuration of MoS$_2$ sheet. Red colored atoms are used to explain the changes in bond length. Dashed lines represent the Mo layer and dash-dotted lines indicate the S layers. Atoms X and Y in (a) possess bond lengths of $2.42~\text{\AA}$ with neighbor sulfur atoms. The bond length between atoms Y-A and Y-D is $2.42~\text{\AA}$ and $2.41~\text{\AA}$, respectively. Atom Y has bond length of $2.26~\text{\AA}$ with atoms B and C. The bond length between atoms Y-E and Y-F is $2.76~\text{\AA}$. Only left portion of the atomic system was shown here.} 
	\label{fig:mos2}
\end{figure}

As we now elaborate, the mechanism enabling the large polarization, and thus large flexoelectric constant in MoS$_{2}$, is different from the other 2D materials.  As shown in Fig.~\ref{fig:mos2}(a), MoS$_{2}$ is a tri-layer 2D materials in which each central Mo atom bonds with the S atoms in the layers above and below.  The thickness of this sheet is defined as the sum of the vertical separation between these layers.  The imposed bending deformation causes the top and bottom S layers to deform differently with respect to the central Mo atom.  For the initial (flat) configuration in Fig.~\ref{fig:mos2}(a), the central Mo atoms, labeled as X and Y, are located  $2.42~\text{\AA}$ away from both the neighboring top and bottom layer S atoms.  This initial atomic configuration also induces non-zero dipole moments to each atom since the $z-$component of $\mathbf{r}_{ij}$ is non-zero. An equal and opposite dipole moment is observed for the top and bottom S atoms due to the equidistant separation with the central Mo atoms, whereas no dipole moment is found on the Mo atoms due to symmetry.


However, after bending, there are significant changes in bond length, as shown in Fig.~\ref{fig:mos2}(b).  The bond lengths between atom X and its nearest S atom neighbors are unchanged even after deformation; the bond lengths Y-A and Y-D are measured as $2.42$ and $2.41~\text{\AA}$.  In contrast, significant changes in bond length result for other nearest S neighbors, where a compression in the Y-B and Y-C bond lengths is identified ($2.42$ to $2.26~\text{\AA}$) in Fig.~\ref{fig:mos2}(b), and where an elongation of the Y-E and Y-F bond lengths ($2.42$ to $2.75~\text{\AA}$) is seen. The identified differences in bond lengths break the symmetry seen in undeformed MoS$_{2}$ in Fig.~\ref{fig:mos2}(a), which leads to non-zero dipole moments, and increases the $E^{q,z}$ contribution to the total electric field as compared to buckled silicene.  

Interestingly, the polarizability of silicene is significantly larger than MoS$_{2}$, i.e. $R_{\text{MoS}_2}/R_{\text{si}}$ is about 0.5, according to Table~\ref{table:charge_width_detail}.  This indicates that MoS$_{2}$ has a significantly higher polarization and flexoelectric constant than buckled silicene for other reasons, starting with the enhanced $\pi-\sigma$ interactions.  Furthermore, a recent DFT study on the bonding characteristics and charge transfer in MoS$_{2}$~\cite{Pike2017} found that the S atoms share their electrons with the Mo atoms, which results in the transfer of electrons back to the Mo atoms.  This charge transfer, coupled with the bond length asymmetry due to bending, results in a large $E^{q,z}$, and thus large dipole moments. 

The flexoelectric coefficients for other members of TMDC group are smaller than MoS$_2$ as shown in Fig.~\ref{fig:flexo_coef_summary}, where the flexoelectric coefficient of WS$_2$ is 3 times smaller than MoS$_2$, and where CrS$_{2}$ has an even smaller value.  We found that the bond length asymmetry between the layers after bending is highest for MoS$_2$ and decreases for WS$_2$ and CrS$_2$, and also that the local difference in radius of curvature for MoS$_2$, WS$_2$ and CrS$_2$ materials is 49, 40 and 26$\%$, respectively, both of which lead to a decreasing contribution of $E^{q,z}$ for WS$_{2}$ and CrS$_{2}$.  The DFT study of Pike \emph{et al.}~\cite{Pike2017} also found a smaller Born effective charge for WS$_2$ compared to MoS$_2$, which supports the observation of lower $E^{q,z}$ for WS$_2$ compared to MoS$_2$.

We also calculated the in-plane flexoelectric constants for all the 2D materials, as summarized in \textcolor{blue}{Appendix}~\ref{sec:in-plane}.  However, we focus on the in-plane flexoelectric constant for MoS$_{2}$ as they are larger than the out-of-plane constants for the other 2D materials.  From Fig.~\ref{fig:PKAniso} and Fig.\ref{fig:Fx_Fz_x}, we observe that the polarization $P^y$ is about an order of magnitude higher than $P^z$ for MoS$_{2}$, whereas the electric field $E^{q,y}$ is less than $E^{q,z}$ for MoS$_2$ ($E^{q,z}/E^{q,y}=7$).   This is because of cancellations in the induced dipole moments in calculating the polarization.  Specifically, the dipole moments $p^y$ have the same sign for all S atoms, whereas $p^z$ has a different sign for the top and bottom planes of S atoms, which induces cancellation of polarization in the z-direction making $P^z$ smaller than $P^y$.  Thus, while $P^y$ is higher than $P^z$, the flexoelectric coefficient $\mu^{yxzx}$ is less than $\mu^{zxzx}$ due to its correlation with $E^{q}$.  Overall, the enhanced $\pi-\sigma$ interactions and bond length asymmetry also leads to strong in-plane electromechanical coupling, and an in-plane flexoelectric constant of $\mu^{yxzx}=0.00962$ nC/m.


An interesting observation from Fig.~\ref{fig:flexo_coef_summary} is that the flexoelectric constants of graphene (C1) and CrS$_{2}$ are nearly equal.  Though CrS$_2$ exhibits higher atomic polarizability and bond length asymmetry, the final dipolar polarization is similar to graphene.  This is because there is a relatively low asymmetry in dipole moment between S atoms in the top and bottom layers in CrS$_{2}$, which results in some cancellation of the induced polarization, leading to a flexoelectric constant that is similar to graphene.  However, for WS$_2$ and MoS$_2$, the increased asymmetry between layers avoids the dipole moment cancellation to achieve larger flexoelectric coefficients.

We find that MoS$_{2}$ has an intrinsic bending flexoelectric constant of $0.032$ nC/m.  This value is about ten times larger than found in graphene, and about 3-5 times larger than seen in the buckled monolayers.  We can compare our computed value with one extracted from the recent experimental study on the electromechanical properties of MoS$_{2}$ reported by Brennan and co-workers~\cite{Brennan2017out}.  In that work, the 
out-of-plane piezoelectric coefficient $(d)$ of MoS$_2$ using piezoresponse force microscopy was measured to be $1.03$ pm/V \cite{Brennan2017out}.  That work also established a relationship between the flexoelectric constant $(\mu)$ and piezoelectric $(d)$ coefficient under the assumption of small length scales and linear electric field \cite{Brennan2017out} as
\begin{equation}
\mu = d Y \frac{t}{2},
\label{eq:mu-exp}
\end{equation}
where $Y$ is the elastic modulus of MoS$_2$ and $t$ is the monolayer thickness of MoS$_2$. With $Y=270$ GPa and $t=0.65$ nm, $\mu$ is about $0.091$ nC/m, which is significantly higher than our calculated value of $0.032$ nC/m.  This difference is due to the usage of elastic modulus in Eq.~(\ref{eq:mu-exp}), where the usage of the bending modulus may be more appropriate.  The bending modulus of MoS$_{2}$ was previously found to be about $9.61$ eV or $65.01$ GPa \cite{Jiang2013}. 
Using this bending modulus, the flexoelectric coefficient from Eq.~\ref{eq:mu-exp} with bending modulus gives a value of $0.021$ nC/m. This value is close compared to the calculated value of $0.032$ nC/m from this work, and demonstrates that the flexoelectric constant for MoS$_{2}$ estimated using the atomistic CD model is in good agreement with experimental measurements. 

\section{Conclusion}
In this work, we used classical atomistic simulations accounting for charge-dipole interactions to study the bending flexoelectric constants for four groups of 2D materials:  graphene allotropes, nitrides, graphene analog monolayer group-IV elements, and TMDs.  Our proposed bending simulations enabled us to directly estimate the flexoelectric constants by eliminating the piezoelectric contribution to the polarization.  In doing so, we were able to analyze the mechanisms underpinning the calculated flexoelectric constants by interpreting them through the electric fields generated from dipole-dipole ($\sigma-\sigma$ bonding) and charge-dipole ($\pi-\sigma$ bonding) interactions.  While the charge-dipole interactions increase with bending curvature, their relative weakness in the flat monolayers (graphene, h-BN) lead to lower flexoelectric constants for these materials.  In contrast, we found that buckling, which occurs in the monolayer group-IV elements, lead to $>$ 10\% increases in flexoelectric constant.  Finally, due to significantly enhanced charge transfer coupled with structural asymmetry due to bending, the TMDCs are found to have the largest flexoelectric constants, including MoS$_{2}$ having a flexoelectric constant ten times larger than graphene.

\appendix

\section{Charge-dipole potential model}
\label{sec:Charge-diople potential model}
\noindent The charge-dipole potential model was first proposed by Olson \textit{et. al.} \cite{olson1978atom}. This model assumes that atom $i$ in a system is associated with a net point charge $q_i$ and a dipole moment $\mathbf{p}_i$. This model has been further developed to overcome the numerical divergence under point charge approximation  \cite{Mayer2007,Mayer2005}. The total electrostatic energy $(E^{\text{CD}})$ for a $N$ atom system is given as
\begin{widetext}
	\begin{equation} 
	\begin{split}
	E^{\text{CD}} & = \frac{1}{2} \sum_i^N \sum_{j,i \neq j}^N q_i T_{ij}^{q-q} q_j -  \sum_i^N \sum_{j,i \neq j}^N q_i \mathbf{T}_{ij}^{q-p} \mathbf{p}_j - \frac{1}{2} \sum_i^N \sum_{j,i \neq j}^N  \mathbf{p}_i \mathbf{T}_{ij}^{p-p} \mathbf{p}_j \\
	& \quad+ \frac{1}{2} \sum_i^N q_i T_{ii}^{q-q} q_i  + \frac{1}{2} \sum_i^N \mathbf{p}_i \mathbf{T}_{ii}^{p-p} \mathbf{p}_i  - \sum_i^N q_i \chi_i - \sum_i^N \mathbf{p}_i \mathbf{E}^{\text{ext}}(\mathbf{r}_i)
	\end{split}
	\label{eq:Etotal}
	\end{equation}
\end{widetext}
where $\chi_i$ is the electron affinity of atom $i$. $T^{q-q}$, $\mathbf{T}^{q-p}$ and $\mathbf{T}^{p-p}$ represent charge-charge, charge-dipole, dipole-dipole interaction coefficients, respectively. The first three terms in Eq.\ref{eq:Etotal} describe the mutual interaction of atomic charges and dipoles among different atoms. The fourth and fifth terms in Eq.~\ref{eq:Etotal} represent the energy required to create a charge and dipole on atom $i$. The sixth term represents the nucleus-to-electron interaction energy. The last term represent the energy due to external electric fields.  $T^{q-q}$ represents the Coulombic interaction between atomic charges which are separated by a distance $r_{ij}$. This coefficient diverges under the point charge approximation when accounting for the self charge term in Eq.~\ref{eq:Etotal}. In order to avoid the divergence of the self energy term, the point charge approximation is modified into atoms with Gaussian distributed charges \cite{Mayer2007,Mayer2005}. The charge distribution for atom $i$ at position $\mathbf{r}$ is
\begin{equation}
\rho_i(\mathbf{r}) = \frac{q_i}{\pi^{3/2}R^3} exp(-\frac{\lvert \mathbf{r} - \mathbf{r}_i \rvert}{R^2})
\end{equation}
where $\mathbf{r}_i$ is position vector of atom $i$. $R$ is equal to $\sqrt{R_{A,i}^2 +  R_{B,j}^2}/\sqrt{2}$, where $R_{A,i}$ represents the width of Gaussian distribution for atom index $i$ with type $A$. $R_{B,j}$ represent the Gaussian distributed charge width for atom type $B$ and with index $j$. For further details about charge-dipole potential refer to \cite{Mayer2007,Mayer2005} and references therein. The parameter $R$ can be estimated from the atomic polarizability. The methodological details are given in next subsection. \\
\noindent For a given atomic configuration, the charge $q$ and dipole moment $\mathbf{p}$ for each atom can be found from the minimization of total electrostatic energy Eq.\ref{eq:Etotal}. The energy minimization with respect to $q_i$ yields
\begin{equation}
T_{ii}^{q-q} q_i + \sum_{j,i\neq j}^{N} T_{ij}^{q-q} q_j - \sum_{j,i\neq j}^{N} \mathbf{T}_{ij}^{q-p} \mathbf{p}_j =  \chi_i.
\label{eq:GDEq}
\end{equation}
The energy minimization with respect to $\mathbf{p}_i$ is
\begin{equation}
\mathbf{T}_{ii}^{p-p} \mathbf{p}_i - \sum_{j,i\neq j}^{N} \mathbf{T}_{ij}^{p-p} \mathbf{p}_j - \sum_{j,i\neq j}^{N} \mathbf{T}_{ij}^{q-p} q_j = \mathbf{E}^{\text{ext}}(\mathbf{r}_i).
\label{eq:GDEp}
\end{equation} 
\noindent From the numerical solution of Eqs. \ref{eq:GDEq} and \ref{eq:GDEp}, the charge and dipole moment are known for the atomic configuration. The electrostatic force and energy calculated from the charge and dipole moment of each atom are supplied to the atomic dynamical equation of motion in addition to the strong short range interactions. Further details on the numerical implementation of charge-dipole potential can be found in recent work by the authors \cite{javvaji2018generation}.  \\
\noindent From the known values of the dipole moment, polarization for the unit cell is defined as the sum of dipole moments of atoms present in that unit cell divided by the volume of the unit cell. The polarization of the $m^{\text{th}}$ unit cell $(\mathbf{P}_m)$ is
\begin{equation}
\mathbf{P}_m = \frac{1}{V_m} \left(\sum_{i=1}^{n} \mathbf{p}_i\right),
\label{eq:Pucell}
\end{equation}
where $n$ is the number of basis atoms present in unit cell $m$, $V_m$ is the volume of the unit cell, and the total polarization is the average among all unit cells in the system. \\ 
\section{Estimation of charge-dipole potential parameter R}
\label{sec:Est-R}
\noindent Consider Eq.~\ref{eq:GDEp} for dipole moments, which is rewritten as
\begin{equation}
\mathbf{T}_{ii}^{p-p} \mathbf{p}_i - \sum_{j,i\neq j}^{N} \mathbf{T}_{ij}^{p-p} \mathbf{p}_j =  \sum_{j,i\neq j}^{N} \mathbf{T}_{ij}^{q-p} q_j + \mathbf{E}^{\text{ext}}(\mathbf{r}_i) .
\label{eq:GDEp2}
\end{equation} 
This represents that the dipole moment of an atom is defined by three different parts:  electric field at position $\mathbf{r}_i$ due to neighboring (i) dipoles $\mathbf{p}_j$ (left hand side second term in Eq.~\ref{eq:GDEp2}); (ii) charges $q_j$ (right hand side first term in Eq.~\ref{eq:GDEp2}) and (iii) from the externally applied electric fields. The diagonal coefficient  $\mathbf{T}_{ii}^{p-p}$ is known as the inverse of atomic polarizability tensor $(\bm{\alpha})$. The mathematical expression for $\mathbf{T}_{ii}^{p-p}$ is given under the CD potential approximations is \cite{Mayer2007,Mayer2005}
\begin{equation}
\mathbf{T}_{ii}^{p-p} = \frac{1}{4\pi\epsilon_0} \frac{\sqrt{2}}{3\sqrt{\pi}R^3} = \frac{1}{\bm{\alpha}_i}
\label{eq:Tpp}
\end{equation}
where $\epsilon_0$ is the dielectric permitivity of vacuum. The CD parameter $R$ is related to the polarizability $\bm{\alpha}$. For an $N$ atomic system, Eq.~\ref{eq:GDEp2} modifies into a matrix-vector system, which is
\begin{equation}
\mathbf{A} \mathbf{p} = \mathbf{E},
\label{eq:vector_est_dipole_alpha}
\end{equation}
where 
\begin{equation}
\mathbf{A} =  
\begin{bmatrix}
\bm{\alpha}_1^{-1} &  \mathbf{T}_{12}^{p-p} & \cdots & \mathbf{T}_{1N}^{p-p}\\
\mathbf{T}_{21}^{p-p} &  \bm{\alpha}_2^{-1} & \cdots & \mathbf{T}_{2N}^{p-p}\\
\cdot &  \cdot & \cdots & \cdot\\
\mathbf{T}_{N1}^{p-p} &  \mathbf{T}_{N2}^{p-p}  & \cdots & \bm{\alpha}_N^{-1}\\
\end{bmatrix}
\label{eq:matrix_est_dipole_alpha}
\end{equation}
and $\mathbf{p}$, $\mathbf{E}$ represent the vector of dipole moments and associated external electric field of each atom, respectively. In order to estimate the polarizability, assuming that the dipoles are experiencing a uniform electric field $(\mathbf{E})$ (which includes both external fields and charge related fields) \cite{silberstein1917molecular,thole1981molecular}, the total dipole moment ($\mathbf{p}_{\text{total}}$) of atomic system is written as
\begin{equation}
\mathbf{p}_{\text{total}} = \bm{\alpha}_{\text{total}} \mathbf{E},
\end{equation}
where $\bm{\alpha}_{\text{total}}$ is the total polarizability of the atomic system, which is expressed as
\begin{equation}
\bm{\alpha}_{\text{total}} = \sum_{i}^N \sum_{j}^N B_{ij},
\label{eq:alphamol}
\end{equation}
where $B_{ij}$ is the components of matrix $\mathbf{A}^{-1}$. Eq.~\ref{eq:alphamol} represents that, in order to estimate $R$, $\bm{\alpha}_{\text{total}}$ has to be known. The polarizability can be calculated from the changes in electronic wave functions. We have used the function $\mathit{polar}$ in Gaussian \cite{frisch2016gaussian} software to estimate $\bm{\alpha}_{\text{total}}$, where details about computing polarizability in DFT calculations  are found elsewhere   \cite{Olsen1985,Sekino1986}. \\
\noindent In order to estimate the $R$ value for graphene, DFT simulations are performed for different sized graphene systems. The isotropic polarizability values from DFT $(\alpha_{\text{total}}^{DFT})$ for these systems are noted. With the atomic coordinate information and assuming $R$ between $0.1$ to $1.0~\text{\AA}$, the total polarizability is calculated $(\alpha_{\text{total}}^{CAL})$ from Eq.~\ref{eq:alphamol}. The atoms present in graphene unit cell are named as A and B. Since both are carbon atoms, it is assumed that $R_A$ is equal to $R_B$ because of the identical electron negativities of these atoms in the unit cell. For graphene samples with size greater than $1$ nm show that for $R=0.64~\text{\AA}$, $\alpha_{\text{total}}^{CAL}$ is identical to $\alpha_{\text{total}}^{DFT}$. The estimated $R$ value for graphene is in close agreement with the estimate based on fullerene structure\cite{Mayer2005}. The size based studies are carried for BN. $R_A$ defines the parameter for N atom and $R_B$ for the B atom. The electron negativities  suggest B atom should have low Gaussian distribution of electronic density when compared to N atom. During the estimation of $\alpha_{\text{total}}^{CAL}$ using Eq.~\ref{eq:alphamol}, it is assumed that $R_A$ is greater than $R_B$ and the estimate matches with the $\alpha_{\text{total}}^{DFT}$ at $0.76$ and $0.35$ for N and B, respectively. Similar type of studies are performed to estimate the CD parameter for other materials. The calculated total polarizability from DFT and derived estimate from Eq.~\ref{eq:alphamol} and the parameter $R$ are tabulated in Table~\ref{table:charge_width_detail}. 
\setlength{\tabcolsep}{2pt}
\begin{table*}
	\begin{ruledtabular}
		\centering
		\caption{Calculation details for each material. The unitcell dimensions $a,b,c$ and $h$ are given in \AA. The bonding interactions are modeled using different types of 'short-range potentials'. $\alpha_{\text{total}}^{DFT}$ and $\alpha_{\text{total}}^{CAL}$ are the polarizbility estimates from DFT and calculated using Eq.~\ref{eq:alphamol} in \AA$^3$. Calculated $R_A$ and $R_B$ in \AA~units, are the CD potential parameter for atom types $A$ and $B$ in the given unitcell. }
		\label{table:charge_width_detail}
		\begin{tabular}{lllllp{2.5cm}cccc}
			material & a & b & c & h & short-range potential & $\alpha_{\text{total}}^{DFT}$ &  $\alpha_{\text{total}}^{CAL}$ & $R_A$ & $R_B$ \\
			\hline
			C1 & 2.46\footnote{\label{a}Reference \cite{enyashin2011graphene}} & 4.26\footref{a} & 3.5\footnote{\label{b}Reference \cite{Ishigami2007}} & 0.0 & AIREBO\footnote{\label{c}Reference \cite{brenner2002second}} & 2.49 & 2.78 & 0.64 & 0.64 \\
			C2 & 4.87\footref{a} & 8.84\footref{a} & 3.5 & 0.0 & AIREBO\footref{c} & 2.46 & 2.72 & 0.64 & 0.64 \\
			C3 & 5.70\footref{a} & 7.56\footref{a} & 3.5 & 0.0 & AIREBO\footref{c} & 2.45 & 2.77 & 0.64 & 0.64 \\
			\hline
			BN & 2.50\footnote{\label{d}Reference \cite{verma2007elastic}} & 2.50\footref{d} & 3.33\footref{d} & 0.0 & Tersoff\footnote{\label{e}Reference \cite{abadi2018fabrication}} & 2.85 & 2.84 & 0.76 & 0.35 \\
			AlN & 3.13\footnote{\label{f}Reference \cite{zhao2016probing}} & 3.13\footref{f} & 3.39\footref{f} & 0.0 & Tersoff\footref{f} & 19.79 & 19.97 & 1.04 & 0.48 \\
			GaN & 3.21\footnote{\label{g}Reference \cite{onen2016gan}}& 3.21\footref{g} & 3.63\footref{g} & 0.0 & Tersoff\footnote{\label{h}Reference \cite{nord2003modelling}} & 15.80 & 16.38 & 1.05 & 0.48 \\
			\hline		
			Si & 3.82\footnote{\label{i}Reference \cite{dimoulas2015silicene}} & 6.62\footref{i} & 2.41\footnote{\label{j}Reference \cite{padilha2015free}} & 0.44\footref{i} & Tersoff\footnote{\label{k}Reference \cite{stillinger1985computer}} & 20.62 & 20.92 & 1.37 & 1.37 \\
			Ge & 3.97\footref{i} & 6.87\footref{i} & 3.20\footnote{\label{l}Reference \cite{Davila2016few}} & 0.65\footref{i} & Tersoff\footnote{\label{m}Reference \cite{mahdizadeh2017optimized}} & 13.43 & 13.14 & 1.27 & 1.27 \\
			Sn & 4.67\footnote{\label{n}Reference \cite{chen2016electronic}} & 8.09\footref{n} & 3.30\footnote{\label{o}Reference \cite{saxena2016stanene}} & 0.89\footref{n} & Tersoff\footref{o} & 15.25 & 15.86 & 1.52 & 1.52 \\
			\hline 
			MoS$_2$ & 3.16\footnote{\label{p}Reference \cite{stewart2013atomistic}} & 3.16\footref{p} & 12.29\footnote{\label{q}Reference \cite{jiang2016first}} & 1.58 & SW\footnote{Reference \cite{jiang2013molecular}} & 12.32 & 12.35 & 0.69 & 1.04 \\
			WS$_2$ & 3.18\footnote{\label{r}Reference \cite{wang2018electronic}} & 3.18\footref{r} & 12.16\footref{r} & 1.56 & SW\footnote{\label{s}Reference \cite{jiang2017parameterization}} & 15.36 & 15.38 & 0.70 & 1.09 \\
			CrS$_2$ & 3.04\footref{q} & 3.04\footref{q} & 14.41\footref{q} & 1.45 & SW\footref{s} & 10.86 & 10.87 & 0.75 & 1.00 \\
		\end{tabular}	
	\end{ruledtabular}	
\end{table*}

\section{Validation of charge-dipole model}
\label{sec:Validation_CD}
\begin{figure}
	\centering
	\includegraphics[width=1.0\linewidth]{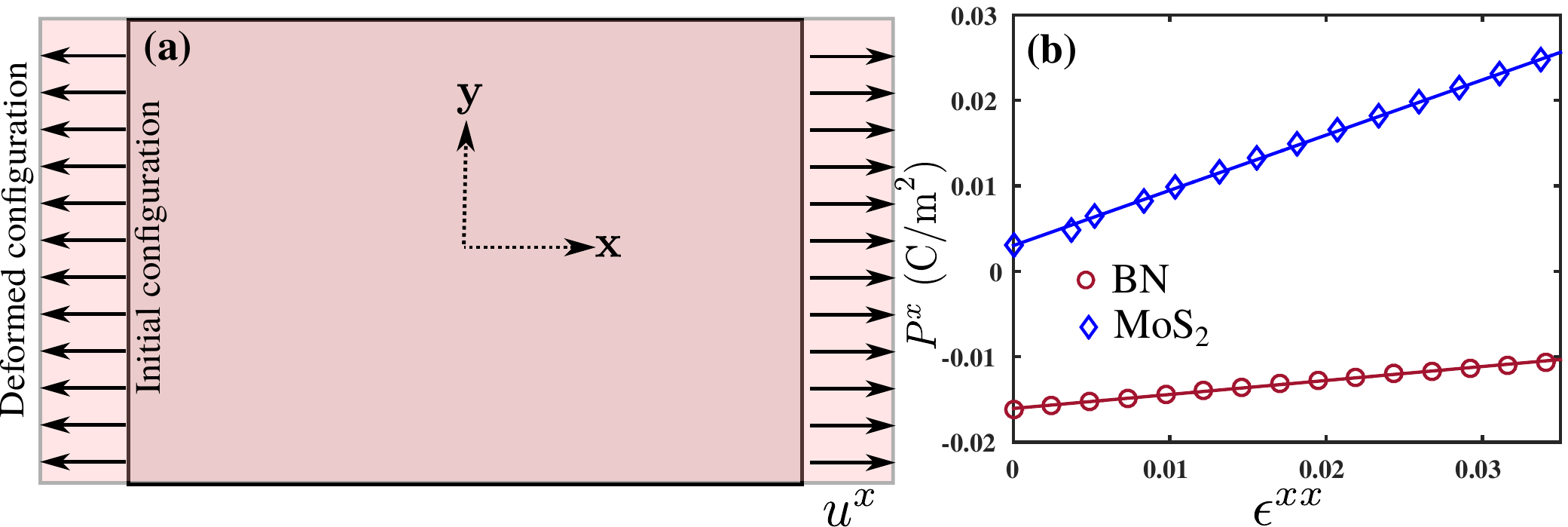}
	\caption{(a) Loading scheme for estimating piezoelectric coefficient and (b) polarization $P^x$ vs strain $\epsilon^{xx}$ for BN and MoS$_2$ material systems.}
	\label{fig:piezoBNMOS2}
\end{figure}

\noindent In this section, the CD model parameters are validated by calculating the piezoelectric coefficients of BN and MoS$_{2}$. A $80~\text{\AA}$ square sheet of BN and MoS$_2$ is subjected to in-plane stretching by displacing $(u^x)$ the left and right ends of the sheet, shown in Fig.~\ref{fig:piezoBNMOS2}(a). The deformed atomic configurations are energy minimized using the conjugate gradient scheme, after which the charge and dipole moments are obtained from Eqs.~\ref{eq:GDEq} and \ref{eq:GDEp} of CD model. A linear variation is observed between total polarization and strain, as shown in Fig.~\ref{fig:piezoBNMOS2}(b). The piezoelectric coefficient (slope of this variation) for BN and MoS$_2$ are in good agreement with the reported DFT estimations (see Table \ref{table:piezo_validate}). This validates the CD parameters derived from DFT and the prediction of electro-mechanical behavior. 
\begin{table}[h]
	\centering
	\caption{Piezoelectric coefficients (C/m$^2$) for BN and MoS$_2$}
	\label{table:piezo_validate}
	\begin{tabular}{c  c c }
		\hline
		Material & Calculated & Reported \\
		\hline
		BN & 0.163 & 0.390\footnote{Reference \cite{Droth2016}}; 0.417\footnote{Reference \cite{blonsky2015ab}}  \\
		MoS$_2$  & 0.646 & 0.564 \footnote{Reference \cite{blonsky2015ab}}; 0.453 \footnote{Reference \cite{Zhu2015}}  \\
		\hline 
	\end{tabular}	
\end{table}
\section{Flexoelectricity in unstable structure}
\begin{figure}
	\centering
	\includegraphics[width=0.8\linewidth]{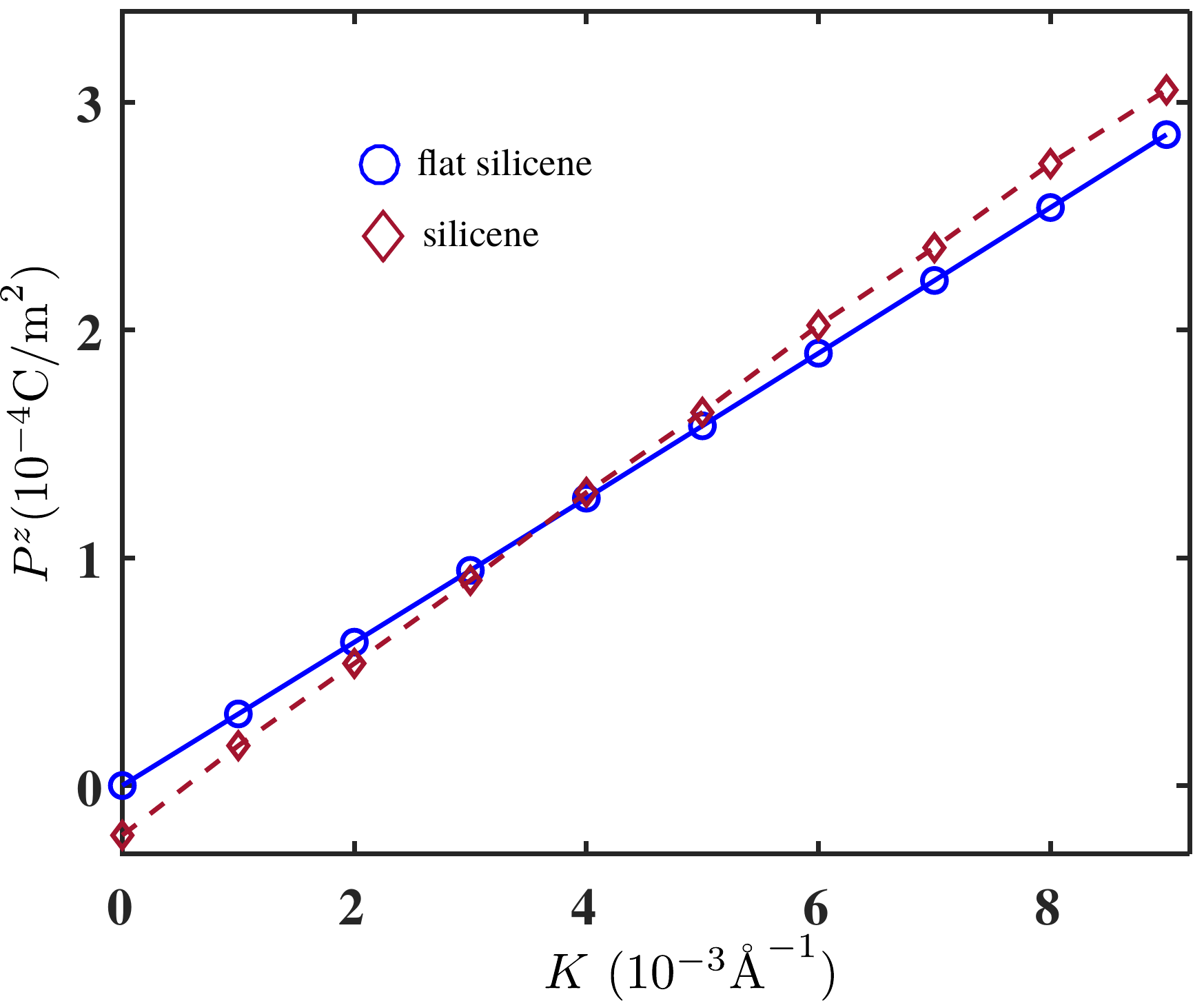}
	\caption{ Polarization $P^z$ vs given strain gradient $K$ for silicene and flat silicene. }
	\label{fig:si-suppl}
\end{figure}

The developed simulation scheme with CD model is applied to unstable 2D structure of silicene. For unstable silicene (flat silicene), the buckling height is assumed as zero to understand the effect of buckling height on the flexoelectric polarization. The polarization response with the strain gradient is given in Fig.~\ref{fig:si-suppl}. The response of silicene was also added for comparison purpose. 
The slope of silicene and flat-silicene are differ only by $15\%$, which is directly connected with the absence of buckling height. 

\section{In-plane flexoelectric polarization} 
\label{sec:in-plane}
\begin{figure}[h]
	\centering
	\includegraphics[scale=0.45]{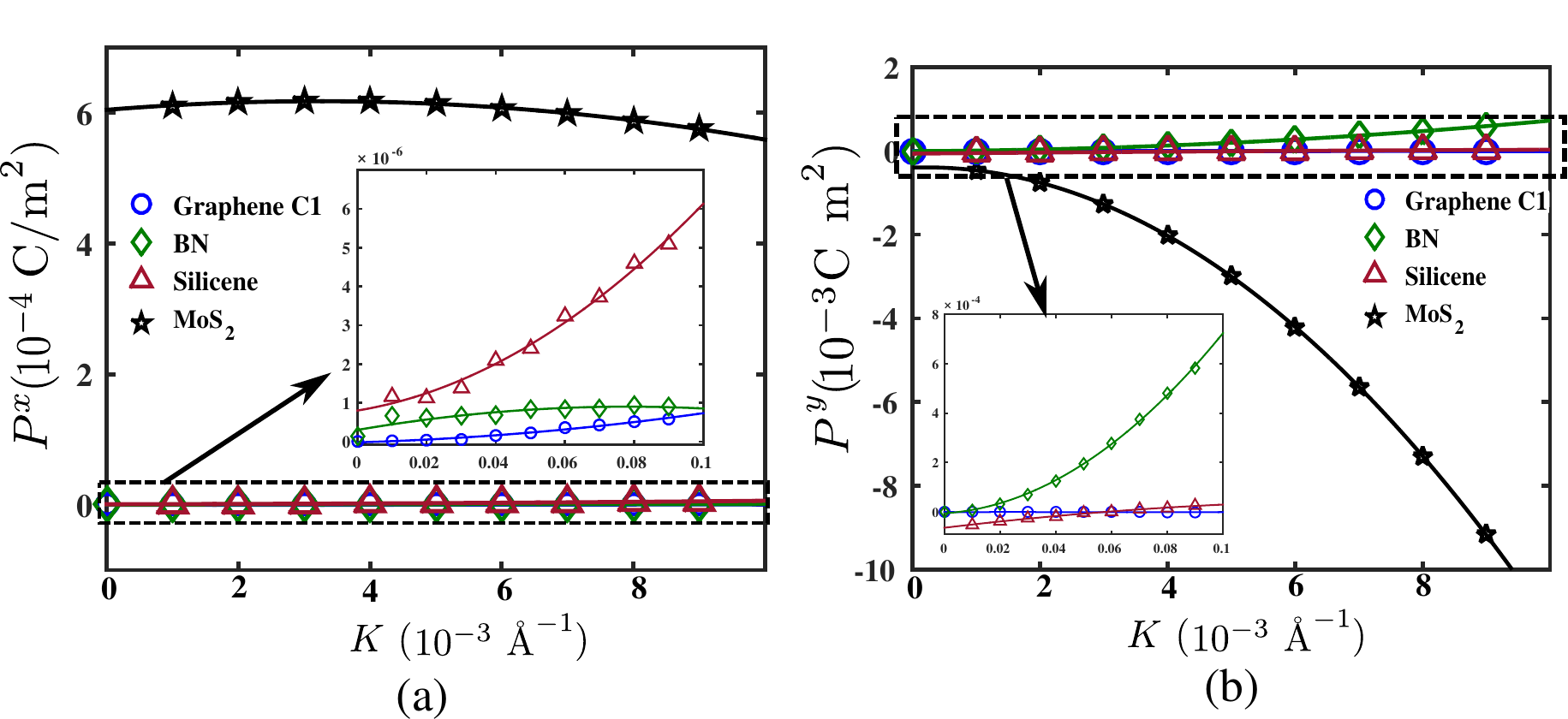}
	\caption{The variation of polarization (a) $P^x$ and (b) $P^y$ with bending curvature $(K)$ for Graphene, BN, Silicene and MoS$_2$ materials. The inset in (a) and (b) represent the polarization variation for materials other than MoS$_2$.}
	\label{fig:PKAniso}
\end{figure}

The variation of in-plane polarization $P^x$ and $P^y$ show a quadratic dependence with bending curvature $K$ as shown in Fig.~\ref{fig:PKAniso}, which is similar to earlier reports for BN \cite{Mele2002,Sai2003,Nakhmanson}. 
\begin{equation}
P^y = d^{yzx} \epsilon^{zx} + \mu^{yxzx} \frac{\partial \epsilon^{zx}}{\partial x} = a_0^y + \textcolor{blue}{\frac{1}{2}}a_1^y K + \textcolor{blue}{\frac{1}{4}} a_2^y K^2
\label{eq:fitPy}
\end{equation}

\begin{equation}
P^x = d^{xzx} \epsilon^{zx} + \mu^{xxzx} \frac{\partial \epsilon^{zx}}{\partial x} = a_0^x + \textcolor{blue}{\frac{1}{2}} a_1^x K + \textcolor{blue}{\frac{1}{4}} a_2^x K^2
\label{eq:fitPx}
\end{equation}
In the above equations, $a_1^x$ and $a_1^y$ have units of C/m, which are those of flexoelectric constants, while $a_0$ and $a_2$ have units of C/m$^2$ and C, respectively.  Taking $a_1$ as the flexoelectric  coefficient, the numerical values for Graphene, BN, Silicene and MoS$_{2}$ are tabulated in Table~\ref{table:muaniso} along with the bending flexoelectric coefficients.  We note that because the mechanical bending we imposed to generate the out-of-plane flexoelectric constants only generate a constant strain in the x-direction from applied deformation in the z-direction, the in-plane polarization that is generated also results in contributions to in-plane piezoelectricity.

\begin{table*}[]
	\begin{ruledtabular}
		\caption{Anisotropic flexoelectric coefficients $(\mu^{zxzx}, \mu^{yxzx} (a_1^y), \mu^{xxzx} (a_1^x))$ given in nC/m. $a_0$ has units of C/m$^2$, while $a_2$ has units of C. }
		\begin{tabular}{llllllll}
			Material & $\mu^{zxzx}$ & $a_1^y (\mu^{yxzx})$ & $a_1^x (\mu^{xxzx})$ &  $a_0^y$ & $a_0^x$ & $a_2^y$ & $a_2^x$ \\
			\hline
			Graphene & 0.00286 & 1.53E-7 & 5.18E-6 & -7.40E-7 & -1.61E-8 & -1.72E-4 & 1.97E-4  \\
			Silicene-flat & 0.00634 & 1.04E-5 & 1.91E-5 & 1.69E-5 & 5.35E-8 & -0.00180 & 0.00072 \\
			\hline
			BN	& 0.00026 & 0.00146 & 3.06E-5 & -5.76E-6 & 3.04E-7 & 0.26296 & -1.96E-4  \\
			Silicene & 0.00728  &  0.0027 & 2.94E-5 & -6.43E-5 & 8.03E-7 & -0.0164  & 0.00152 \\
			MoS$_2$ & 0.03194 & 0.00962 & 0.00164 & -0.00039 & 0.00060 & -4.5484 & -0.0512
			\label{table:muaniso}
		\end{tabular}
	\end{ruledtabular}
\end{table*}

From Table.~\ref{table:muaniso}, the in-plane flexoelectric coefficients $(\mu^{yxzx}~\text{and}~\mu^{xxzx})$ are significantly smaller than the out-of-plane coefficient $(\mu^{zxzx})$ of graphene under bending deformation. The corresponding in-plane polarizations $P^x$ and $P^y$ are also lower than the out of plane polarization $P^z$. This implies that the in-plane $\pi-\sigma$ interactions for graphene generate relatively small in-plane dipole moments. \textcolor{blue}{A symmetry analysis can be used to show that $\mu^{yxzx}~\text{and}~\mu^{xxzx}$ are zero \cite{Shu2011}, which implies that the graphene system is isotropic \cite{Ahmadpoor2015}. } Similarly, lower in-plane flexoelectric coefficients are observed for flat silicene (see Table.~\ref{table:muaniso}). 

In the case of  \textcolor{blue}{anisotropic} BN \cite{Ahmadpoor2015}, the in-plane coefficient $(\mu^{yxzx})$ is nearly 5 times larger than the out-of-plane coefficient $(\mu^{zxzx})$, as shown in Table.~\ref{table:muaniso}, while Figs.~\ref{fig:PKAniso}(b) and Fig.~\ref{fig:Fx_Fz_x} show that the polarization $P^y$ is higher than $P^z$.  In addition, the charge-dipole coupling induced electric field $E^{q,y}$ is greater than $E^{q,z}$, and that the ratio of $E^{q,y}/E^{q,z}$ is about $4.6$, which is similar to the ratio between the in-plane and out of plane flexoelectric coefficients. First principles calculations for a corrugated BN sheet \cite{Naumov2009} provide significant in-plane polarization, which is related to $\pi$ and $\sigma$ chemical bond shifts due to the out-of-plane atomic displacements. The very small difference in out of plane displacements of the B and N atoms \cite{Wirtz2003,Moon2004} leads to relatively small out-of-plane dipole moments, and also suggests that in-plane $\pi-\sigma$ interactions are stronger which makes $\mu^{yxzx}$ is higher than $\mu^{zxzx}$.

The flexoelectric coefficient $\mu^{zxzx}$ is higher than $\mu^{yxzx}$ for buckled silicene.  The corresponding polarization $P^z$ is greater than $P^y$, as observed from Fig.~\ref{fig:Fx_Fz_x} and \ref{fig:PKAniso}(b), and the out of plane electric field is larger than the in-plane electric fields, which implies that the $\pi-\sigma$ coupling is stronger out-of-plane than in-plane.  It is also noted that the atomic buckling in silicene significantly enhances the in-plane flexoelectric coefficient compared to flat silicene (see values of $\mu^{yxzx}$ for silicene and silicene-flat materials in Table.~\ref{table:muaniso}). Because the buckling height of silicene is significantly larger than seen in BN, larger dipole moments and thus larger in-plane and out-of-plane flexoelectric constants are predicted for buckled silicene as compared to BN.

\begin{figure}[h]
	\centering
	\includegraphics[scale=0.55]{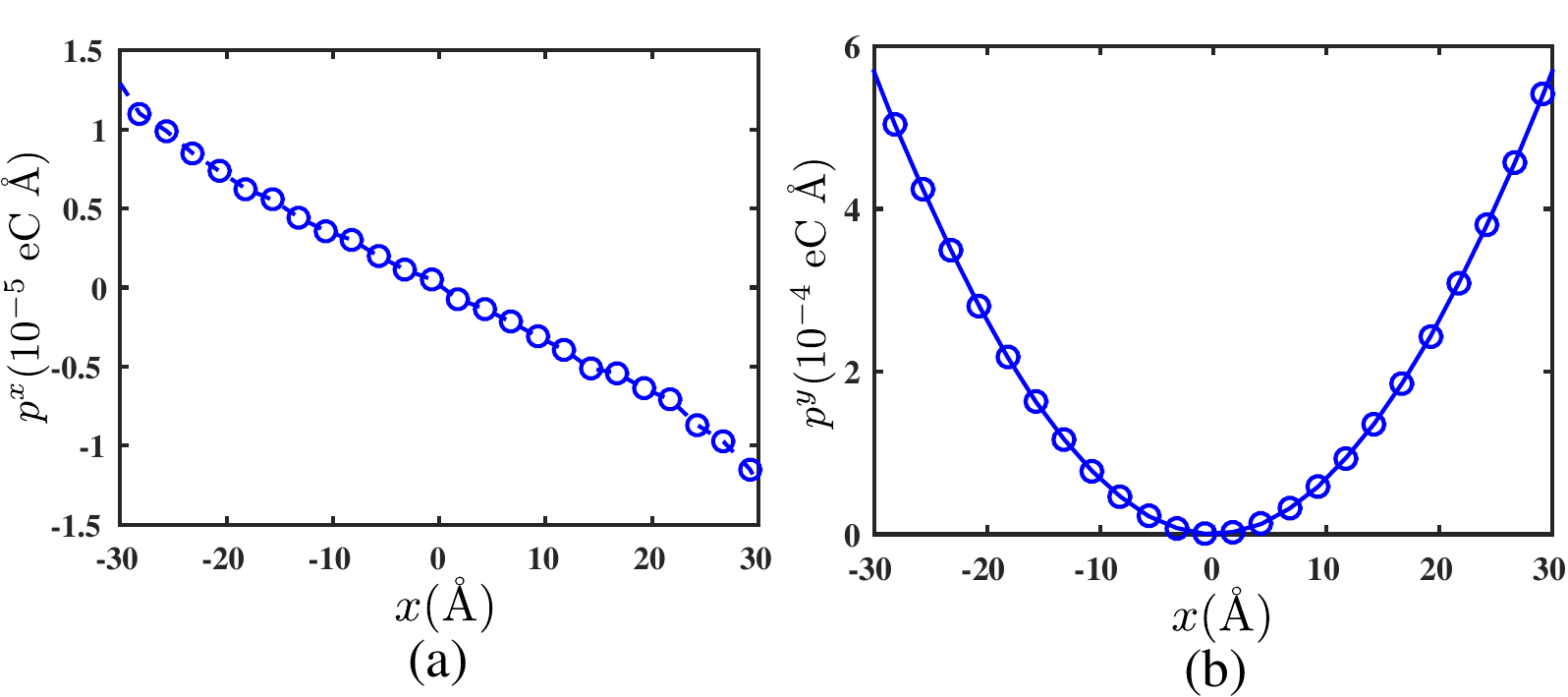}
	\caption{Bin-wise distribution of polarization (a) $P^x$ and (b) $P^y$ along $x$ axis for BN at a curvature of $K=0.08$ nm$^{-1}$.}
	\label{fig:PxPyxbinsBN}
\end{figure}

In BN, silicene and MoS$_2$, it is observed from Fig.~\ref{fig:PKAniso} for BN that the in-plane polarization $P^y$ is higher than the in-plane polarization $P^x$. In the present study, the $x-$direction is considered as the armchair configuration while the $y-$direction represents the zigzag configuration, which means that polarization in the zigzag direction is higher than in the armchair direction, which was previously observed for BN~\cite{Naumov2009}.  To further confirm this, we rotated the atomic system to change the $x-$direction to zigzag and the $y-$direction to armchair and repeated the bending test. There is no change in polarization $P^z$ and the corresponding flexoelectric coefficients. When coming to in-plane polarizations, we found $p^x$ (zigzag) is higher than $p^y$ (armchair).  We also observed that $p^x$ is linear and $p^y$ is parabolic, which is also the same form for the strain fields that were observed in the armchair and zigzag directions.  This implies that the local atomic configuration strongly impacts the deformation, and thus the induced polarization, which was also observed in silicene and MoS$_{2}$. \textcolor{blue}{The observation of anisotropic in-plane polarization due to bending is similar to the earlier findings reviewed by Ahmadpoor and Sharma \cite{Ahmadpoor2015}.}

In summary, we find that monatomic unit cells, such as graphene and flat silicene, do not exhibit spatial variations in the out of plane displacements due to bending, whereas MoS$_{2}$, buckled silicene and BN do exhibit, to varying degrees, spatial variations in the out of plane displacements due to bending.  As a result, the in-plane charge-dipole interactions are week for graphene and flat silicene, resulting in low in-plane flexoelectric constants.  In contrast, MoS$_{2}$ exhibits significant structural asymmetry under bending, which enhances the in-plane $\pi-\sigma$ coupling, with a similar effect seen in buckled silicene.  While BN does not exhibit some spatial variation in the out of plane displacements, the out of plane displacements are relatively small, and as such, the in-plane flexoelectric constants are smaller than buckled silicene and MoS$_{2}$.

\bibliographystyle{apsrev4-1}
\bibliography{reference}

\begin{thebibliography}{78}%
\makeatletter
\providecommand \@ifxundefined [1]{%
 \@ifx{#1\undefined}
}%
\providecommand \@ifnum [1]{%
 \ifnum #1\expandafter \@firstoftwo
 \else \expandafter \@secondoftwo
 \fi
}%
\providecommand \@ifx [1]{%
 \ifx #1\expandafter \@firstoftwo
 \else \expandafter \@secondoftwo
 \fi
}%
\providecommand \natexlab [1]{#1}%
\providecommand \enquote  [1]{``#1''}%
\providecommand \bibnamefont  [1]{#1}%
\providecommand \bibfnamefont [1]{#1}%
\providecommand \citenamefont [1]{#1}%
\providecommand \href@noop [0]{\@secondoftwo}%
\providecommand \href [0]{\begingroup \@sanitize@url \@href}%
\providecommand \@href[1]{\@@startlink{#1}\@@href}%
\providecommand \@@href[1]{\endgroup#1\@@endlink}%
\providecommand \@sanitize@url [0]{\catcode `\\12\catcode `\$12\catcode
  `\&12\catcode `\#12\catcode `\^12\catcode `\_12\catcode `\%12\relax}%
\providecommand \@@startlink[1]{}%
\providecommand \@@endlink[0]{}%
\providecommand \url  [0]{\begingroup\@sanitize@url \@url }%
\providecommand \@url [1]{\endgroup\@href {#1}{\urlprefix }}%
\providecommand \urlprefix  [0]{URL }%
\providecommand \Eprint [0]{\href }%
\providecommand \doibase [0]{http://dx.doi.org/}%
\providecommand \selectlanguage [0]{\@gobble}%
\providecommand \bibinfo  [0]{\@secondoftwo}%
\providecommand \bibfield  [0]{\@secondoftwo}%
\providecommand \translation [1]{[#1]}%
\providecommand \BibitemOpen [0]{}%
\providecommand \bibitemStop [0]{}%
\providecommand \bibitemNoStop [0]{.\EOS\space}%
\providecommand \EOS [0]{\spacefactor3000\relax}%
\providecommand \BibitemShut  [1]{\csname bibitem#1\endcsname}%
\let\auto@bib@innerbib\@empty
\bibitem [{\citenamefont {Ikeda}(1996)}]{ikeda1996fundamentals}%
  \BibitemOpen
  \bibfield  {author} {\bibinfo {author} {\bibfnamefont {T.}~\bibnamefont
  {Ikeda}},\ }\href@noop {} {\emph {\bibinfo {title} {Fundamentals of
  piezoelectricity}}}\ (\bibinfo  {publisher} {Oxford university press},\
  \bibinfo {year} {1996})\BibitemShut {NoStop}%
\bibitem [{\citenamefont {Heywang}\ \emph {et~al.}(2008)\citenamefont
  {Heywang}, \citenamefont {Lubitz},\ and\ \citenamefont
  {Wersing}}]{heywang2008piezoelectricity}%
  \BibitemOpen
  \bibfield  {author} {\bibinfo {author} {\bibfnamefont {W.}~\bibnamefont
  {Heywang}}, \bibinfo {author} {\bibfnamefont {K.}~\bibnamefont {Lubitz}}, \
  and\ \bibinfo {author} {\bibfnamefont {W.}~\bibnamefont {Wersing}},\ }\href
  {\doibase 10.1007/978-3-540-68683-5} {\emph {\bibinfo {title}
  {Piezoelectricity: evolution and future of a technology}}},\ Vol.\ \bibinfo
  {volume} {114}\ (\bibinfo  {publisher} {Springer Science \& Business Media},\
  \bibinfo {year} {2008})\BibitemShut {NoStop}%
\bibitem [{\citenamefont {Ma}\ and\ \citenamefont
  {Cross}(2006)}]{ma2006flexoelectricity}%
  \BibitemOpen
  \bibfield  {author} {\bibinfo {author} {\bibfnamefont {W.}~\bibnamefont
  {Ma}}\ and\ \bibinfo {author} {\bibfnamefont {L.~E.}\ \bibnamefont {Cross}},\
  }\href {\doibase 10.1063/1.2211309} {\bibfield  {journal} {\bibinfo
  {journal} {Appl. Phys. Lett.}\ }\textbf {\bibinfo {volume} {88}},\ \bibinfo
  {pages} {232902} (\bibinfo {year} {2006})}\BibitemShut {NoStop}%
\bibitem [{\citenamefont {Harden}\ \emph {et~al.}(2006)\citenamefont {Harden},
  \citenamefont {Mbanga}, \citenamefont {{\'E}ber}, \citenamefont
  {Fodor-Csorba}, \citenamefont {Sprunt}, \citenamefont {Gleeson},\ and\
  \citenamefont {J{\'a}kli}}]{harden2006giant}%
  \BibitemOpen
  \bibfield  {author} {\bibinfo {author} {\bibfnamefont {J.}~\bibnamefont
  {Harden}}, \bibinfo {author} {\bibfnamefont {B.}~\bibnamefont {Mbanga}},
  \bibinfo {author} {\bibfnamefont {N.}~\bibnamefont {{\'E}ber}}, \bibinfo
  {author} {\bibfnamefont {K.}~\bibnamefont {Fodor-Csorba}}, \bibinfo {author}
  {\bibfnamefont {S.}~\bibnamefont {Sprunt}}, \bibinfo {author} {\bibfnamefont
  {J.~T.}\ \bibnamefont {Gleeson}}, \ and\ \bibinfo {author} {\bibfnamefont
  {A.}~\bibnamefont {J{\'a}kli}},\ }\href {\doibase
  10.1103/PhysRevLett.97.157802} {\bibfield  {journal} {\bibinfo  {journal}
  {Phys. Rev. Lett.}\ }\textbf {\bibinfo {volume} {97}},\ \bibinfo {pages}
  {157802} (\bibinfo {year} {2006})}\BibitemShut {NoStop}%
\bibitem [{\citenamefont {Hong}\ \emph {et~al.}(2010)\citenamefont {Hong},
  \citenamefont {Catalan}, \citenamefont {Scott},\ and\ \citenamefont
  {Artacho}}]{hong2010flexoelectricity}%
  \BibitemOpen
  \bibfield  {author} {\bibinfo {author} {\bibfnamefont {J.}~\bibnamefont
  {Hong}}, \bibinfo {author} {\bibfnamefont {G.}~\bibnamefont {Catalan}},
  \bibinfo {author} {\bibfnamefont {J.}~\bibnamefont {Scott}}, \ and\ \bibinfo
  {author} {\bibfnamefont {E.}~\bibnamefont {Artacho}},\ }\href {\doibase
  10.1088/0953-8984/22/11/112201} {\bibfield  {journal} {\bibinfo  {journal}
  {J. Phys.: Condens. Matter}\ }\textbf {\bibinfo {volume} {22}},\ \bibinfo
  {pages} {112201} (\bibinfo {year} {2010})}\BibitemShut {NoStop}%
\bibitem [{\citenamefont {Kalinin}\ and\ \citenamefont
  {Meunier}(2008)}]{kalinin2008electronic}%
  \BibitemOpen
  \bibfield  {author} {\bibinfo {author} {\bibfnamefont {S.~V.}\ \bibnamefont
  {Kalinin}}\ and\ \bibinfo {author} {\bibfnamefont {V.}~\bibnamefont
  {Meunier}},\ }\href {\doibase 10.1103/PhysRevB.77.033403} {\bibfield
  {journal} {\bibinfo  {journal} {Phys. Rev. B}\ }\textbf {\bibinfo {volume}
  {77}},\ \bibinfo {pages} {033403} (\bibinfo {year} {2008})}\BibitemShut
  {NoStop}%
\bibitem [{\citenamefont {Krichen}\ and\ \citenamefont
  {Sharma}(2016)}]{krichen2016flexoelectricity}%
  \BibitemOpen
  \bibfield  {author} {\bibinfo {author} {\bibfnamefont {S.}~\bibnamefont
  {Krichen}}\ and\ \bibinfo {author} {\bibfnamefont {P.}~\bibnamefont
  {Sharma}},\ }\href {\doibase 10.1115/1.4032378} {\bibfield  {journal}
  {\bibinfo  {journal} {J. of Appl. Mech.}\ }\textbf {\bibinfo {volume} {83}},\
  \bibinfo {pages} {030801} (\bibinfo {year} {2016})}\BibitemShut {NoStop}%
\bibitem [{\citenamefont {Tagantsev}(1986)}]{Tagantsev1986}%
  \BibitemOpen
  \bibfield  {author} {\bibinfo {author} {\bibfnamefont {A.~K.}\ \bibnamefont
  {Tagantsev}},\ }\href {\doibase 10.1103/PhysRevB.34.5883} {\bibfield
  {journal} {\bibinfo  {journal} {Phys. Rev. B}\ }\textbf {\bibinfo {volume}
  {34}},\ \bibinfo {pages} {5883} (\bibinfo {year} {1986})}\BibitemShut
  {NoStop}%
\bibitem [{\citenamefont {Wang}\ \emph {et~al.}(2015)\citenamefont {Wang},
  \citenamefont {Tian}, \citenamefont {Xie}, \citenamefont {Shu}, \citenamefont
  {Mi}, \citenamefont {Mohammad}, \citenamefont {Xie}, \citenamefont {Wang},
  \citenamefont {Xu},\ and\ \citenamefont {Ren}}]{Wang2015observattion}%
  \BibitemOpen
  \bibfield  {author} {\bibinfo {author} {\bibfnamefont {X.}~\bibnamefont
  {Wang}}, \bibinfo {author} {\bibfnamefont {H.}~\bibnamefont {Tian}}, \bibinfo
  {author} {\bibfnamefont {W.}~\bibnamefont {Xie}}, \bibinfo {author}
  {\bibfnamefont {Y.}~\bibnamefont {Shu}}, \bibinfo {author} {\bibfnamefont
  {W.-T.}\ \bibnamefont {Mi}}, \bibinfo {author} {\bibfnamefont {M.~A.}\
  \bibnamefont {Mohammad}}, \bibinfo {author} {\bibfnamefont {Q.-Y.}\
  \bibnamefont {Xie}}, \bibinfo {author} {\bibfnamefont {Y.}~\bibnamefont
  {Wang}}, \bibinfo {author} {\bibfnamefont {J.-B.}\ \bibnamefont {Xu}}, \ and\
  \bibinfo {author} {\bibfnamefont {T.-L.}\ \bibnamefont {Ren}},\ }\href
  {\doibase 10.1038/am.2014.124} {\bibfield  {journal} {\bibinfo  {journal}
  {NPG Asia Materials}\ }\textbf {\bibinfo {volume} {7}},\ \bibinfo {pages}
  {e154} (\bibinfo {year} {2015})}\BibitemShut {NoStop}%
\bibitem [{\citenamefont {Song}\ \emph {et~al.}(2017)\citenamefont {Song},
  \citenamefont {Hui}, \citenamefont {Knobloch}, \citenamefont {Wang},
  \citenamefont {Fan}, \citenamefont {Grasser}, \citenamefont {Jing},
  \citenamefont {Shi},\ and\ \citenamefont {Lanza}}]{Song2017piezo}%
  \BibitemOpen
  \bibfield  {author} {\bibinfo {author} {\bibfnamefont {X.}~\bibnamefont
  {Song}}, \bibinfo {author} {\bibfnamefont {F.}~\bibnamefont {Hui}}, \bibinfo
  {author} {\bibfnamefont {T.}~\bibnamefont {Knobloch}}, \bibinfo {author}
  {\bibfnamefont {B.}~\bibnamefont {Wang}}, \bibinfo {author} {\bibfnamefont
  {Z.}~\bibnamefont {Fan}}, \bibinfo {author} {\bibfnamefont {T.}~\bibnamefont
  {Grasser}}, \bibinfo {author} {\bibfnamefont {X.}~\bibnamefont {Jing}},
  \bibinfo {author} {\bibfnamefont {Y.}~\bibnamefont {Shi}}, \ and\ \bibinfo
  {author} {\bibfnamefont {M.}~\bibnamefont {Lanza}},\ }\href {\doibase
  10.1063/1.5000496} {\bibfield  {journal} {\bibinfo  {journal} {Appl. Phys.
  Lett}\ }\textbf {\bibinfo {volume} {111}},\ \bibinfo {pages} {083107}
  (\bibinfo {year} {2017})}\BibitemShut {NoStop}%
\bibitem [{\citenamefont {Brennan}\ \emph {et~al.}(2017)\citenamefont
  {Brennan}, \citenamefont {Ghosh}, \citenamefont {Koul}, \citenamefont
  {Banerjee}, \citenamefont {Lu},\ and\ \citenamefont {Yu}}]{Brennan2017out}%
  \BibitemOpen
  \bibfield  {author} {\bibinfo {author} {\bibfnamefont {C.~J.}\ \bibnamefont
  {Brennan}}, \bibinfo {author} {\bibfnamefont {R.}~\bibnamefont {Ghosh}},
  \bibinfo {author} {\bibfnamefont {K.}~\bibnamefont {Koul}}, \bibinfo {author}
  {\bibfnamefont {S.~K.}\ \bibnamefont {Banerjee}}, \bibinfo {author}
  {\bibfnamefont {N.}~\bibnamefont {Lu}}, \ and\ \bibinfo {author}
  {\bibfnamefont {E.~T.}\ \bibnamefont {Yu}},\ }\href@noop {} {\bibfield
  {journal} {\bibinfo  {journal} {Nano Lett,}\ }\textbf {\bibinfo {volume}
  {17}},\ \bibinfo {pages} {5464} (\bibinfo {year} {2017})}\BibitemShut
  {NoStop}%
\bibitem [{\citenamefont {Zheng}\ \emph {et~al.}(2017)\citenamefont {Zheng},
  \citenamefont {Lee},\ and\ \citenamefont {Feng}}]{Zheng2017hexagonal}%
  \BibitemOpen
  \bibfield  {author} {\bibinfo {author} {\bibfnamefont {X.-Q.}\ \bibnamefont
  {Zheng}}, \bibinfo {author} {\bibfnamefont {J.}~\bibnamefont {Lee}}, \ and\
  \bibinfo {author} {\bibfnamefont {P.~X.-L.}\ \bibnamefont {Feng}},\ }\href
  {\doibase 10.1038/micronano.2017.38} {\bibfield  {journal} {\bibinfo
  {journal} {Microsystems \& Nanoengineering}\ }\textbf {\bibinfo {volume}
  {3}},\ \bibinfo {pages} {17038} (\bibinfo {year} {2017})}\BibitemShut
  {NoStop}%
\bibitem [{\citenamefont {Zelisko}\ \emph {et~al.}(2014)\citenamefont
  {Zelisko}, \citenamefont {Hanlumyuang}, \citenamefont {Yang}, \citenamefont
  {Liu}, \citenamefont {Lei}, \citenamefont {Li}, \citenamefont {Ajayan},\ and\
  \citenamefont {Sharma}}]{Zelisko2014anomalous}%
  \BibitemOpen
  \bibfield  {author} {\bibinfo {author} {\bibfnamefont {M.}~\bibnamefont
  {Zelisko}}, \bibinfo {author} {\bibfnamefont {Y.}~\bibnamefont
  {Hanlumyuang}}, \bibinfo {author} {\bibfnamefont {S.}~\bibnamefont {Yang}},
  \bibinfo {author} {\bibfnamefont {Y.}~\bibnamefont {Liu}}, \bibinfo {author}
  {\bibfnamefont {C.}~\bibnamefont {Lei}}, \bibinfo {author} {\bibfnamefont
  {J.}~\bibnamefont {Li}}, \bibinfo {author} {\bibfnamefont {P.~M.}\
  \bibnamefont {Ajayan}}, \ and\ \bibinfo {author} {\bibfnamefont
  {P.}~\bibnamefont {Sharma}},\ }\href {\doibase 10.1038/ncomms5284} {\bibfield
   {journal} {\bibinfo  {journal} {Nat. Comm.}\ }\textbf {\bibinfo {volume}
  {5}},\ \bibinfo {pages} {4284} (\bibinfo {year} {2014})}\BibitemShut
  {NoStop}%
\bibitem [{\citenamefont {Blonsky}\ \emph {et~al.}(2015)\citenamefont
  {Blonsky}, \citenamefont {Zhuang}, \citenamefont {Singh},\ and\ \citenamefont
  {Hennig}}]{blonsky2015ab}%
  \BibitemOpen
  \bibfield  {author} {\bibinfo {author} {\bibfnamefont {M.~N.}\ \bibnamefont
  {Blonsky}}, \bibinfo {author} {\bibfnamefont {H.~L.}\ \bibnamefont {Zhuang}},
  \bibinfo {author} {\bibfnamefont {A.~K.}\ \bibnamefont {Singh}}, \ and\
  \bibinfo {author} {\bibfnamefont {R.~G.}\ \bibnamefont {Hennig}},\ }\href
  {\doibase 10.1021/acsnano.5b03394} {\bibfield  {journal} {\bibinfo  {journal}
  {ACS nano}\ }\textbf {\bibinfo {volume} {9}},\ \bibinfo {pages} {9885}
  (\bibinfo {year} {2015})}\BibitemShut {NoStop}%
\bibitem [{\citenamefont {Kundalwal}\ \emph {et~al.}(2017)\citenamefont
  {Kundalwal}, \citenamefont {Meguid},\ and\ \citenamefont
  {Weng}}]{Kundalwal2017}%
  \BibitemOpen
  \bibfield  {author} {\bibinfo {author} {\bibfnamefont {S.~I.}\ \bibnamefont
  {Kundalwal}}, \bibinfo {author} {\bibfnamefont {S.~A.}\ \bibnamefont
  {Meguid}}, \ and\ \bibinfo {author} {\bibfnamefont {G.~J.}\ \bibnamefont
  {Weng}},\ }\href {\doibase 10.1016/j.carbon.2017.03.013} {\bibfield
  {journal} {\bibinfo  {journal} {Carbon}\ }\textbf {\bibinfo {volume} {117}},\
  \bibinfo {pages} {462} (\bibinfo {year} {2017})}\BibitemShut {NoStop}%
\bibitem [{\citenamefont {Chandratre}\ and\ \citenamefont
  {Sharma}(2012)}]{chandratre2012coaxing}%
  \BibitemOpen
  \bibfield  {author} {\bibinfo {author} {\bibfnamefont {S.}~\bibnamefont
  {Chandratre}}\ and\ \bibinfo {author} {\bibfnamefont {P.}~\bibnamefont
  {Sharma}},\ }\href {\doibase 10.1063/1.3676084} {\bibfield  {journal}
  {\bibinfo  {journal} {Applied Physics Letters}\ }\textbf {\bibinfo {volume}
  {100}},\ \bibinfo {pages} {023114} (\bibinfo {year} {2012})}\BibitemShut
  {NoStop}%
\bibitem [{\citenamefont {Duerloo}\ and\ \citenamefont
  {Reed}(2013)}]{duerloo2013flexural}%
  \BibitemOpen
  \bibfield  {author} {\bibinfo {author} {\bibfnamefont {K.-A.~N.}\
  \bibnamefont {Duerloo}}\ and\ \bibinfo {author} {\bibfnamefont {E.~J.}\
  \bibnamefont {Reed}},\ }\href {\doibase 10.1021/nl4001635} {\bibfield
  {journal} {\bibinfo  {journal} {Nano letters}\ }\textbf {\bibinfo {volume}
  {13}},\ \bibinfo {pages} {1681} (\bibinfo {year} {2013})}\BibitemShut
  {NoStop}%
\bibitem [{\citenamefont {Zhang}(2017)}]{zhang2017boron}%
  \BibitemOpen
  \bibfield  {author} {\bibinfo {author} {\bibfnamefont {J.}~\bibnamefont
  {Zhang}},\ }\href {\doibase 10.1016/j.nanoen.2017.10.005} {\bibfield
  {journal} {\bibinfo  {journal} {Nano Energy}\ }\textbf {\bibinfo {volume}
  {41}},\ \bibinfo {pages} {460} (\bibinfo {year} {2017})}\BibitemShut
  {NoStop}%
\bibitem [{\citenamefont {Zhou}\ \emph {et~al.}(2016)\citenamefont {Zhou},
  \citenamefont {Liu}, \citenamefont {Huang}, \citenamefont {Zhang},
  \citenamefont {Zhang},\ and\ \citenamefont {Wang}}]{zhou2016theoretical}%
  \BibitemOpen
  \bibfield  {author} {\bibinfo {author} {\bibfnamefont {Y.}~\bibnamefont
  {Zhou}}, \bibinfo {author} {\bibfnamefont {W.}~\bibnamefont {Liu}}, \bibinfo
  {author} {\bibfnamefont {X.}~\bibnamefont {Huang}}, \bibinfo {author}
  {\bibfnamefont {A.}~\bibnamefont {Zhang}}, \bibinfo {author} {\bibfnamefont
  {Y.}~\bibnamefont {Zhang}}, \ and\ \bibinfo {author} {\bibfnamefont {Z.~L.}\
  \bibnamefont {Wang}},\ }\href@noop {} {\bibfield  {journal} {\bibinfo
  {journal} {Nano Research}\ }\textbf {\bibinfo {volume} {9}},\ \bibinfo
  {pages} {800} (\bibinfo {year} {2016})}\BibitemShut {NoStop}%
\bibitem [{\citenamefont {Javvaji}\ \emph {et~al.}(2018)\citenamefont
  {Javvaji}, \citenamefont {He},\ and\ \citenamefont
  {Zhuang}}]{javvaji2018generation}%
  \BibitemOpen
  \bibfield  {author} {\bibinfo {author} {\bibfnamefont {B.}~\bibnamefont
  {Javvaji}}, \bibinfo {author} {\bibfnamefont {B.}~\bibnamefont {He}}, \ and\
  \bibinfo {author} {\bibfnamefont {X.}~\bibnamefont {Zhuang}},\ }\href
  {\doibase 10.1088/1361-6528/aab5ad} {\bibfield  {journal} {\bibinfo
  {journal} {Nanotechnology}\ }\textbf {\bibinfo {volume} {29}},\ \bibinfo
  {pages} {225702} (\bibinfo {year} {2018})}\BibitemShut {NoStop}%
\bibitem [{\citenamefont {Wu}\ \emph {et~al.}(2014)\citenamefont {Wu},
  \citenamefont {Wang}, \citenamefont {Li}, \citenamefont {Zhang},
  \citenamefont {Lin}, \citenamefont {Niu}, \citenamefont {Chenet},
  \citenamefont {Zhang}, \citenamefont {Hao}, \citenamefont {Heinz} \emph
  {et~al.}}]{wu2014piezoelectricity}%
  \BibitemOpen
  \bibfield  {author} {\bibinfo {author} {\bibfnamefont {W.}~\bibnamefont
  {Wu}}, \bibinfo {author} {\bibfnamefont {L.}~\bibnamefont {Wang}}, \bibinfo
  {author} {\bibfnamefont {Y.}~\bibnamefont {Li}}, \bibinfo {author}
  {\bibfnamefont {F.}~\bibnamefont {Zhang}}, \bibinfo {author} {\bibfnamefont
  {L.}~\bibnamefont {Lin}}, \bibinfo {author} {\bibfnamefont {S.}~\bibnamefont
  {Niu}}, \bibinfo {author} {\bibfnamefont {D.}~\bibnamefont {Chenet}},
  \bibinfo {author} {\bibfnamefont {X.}~\bibnamefont {Zhang}}, \bibinfo
  {author} {\bibfnamefont {Y.}~\bibnamefont {Hao}}, \bibinfo {author}
  {\bibfnamefont {T.~F.}\ \bibnamefont {Heinz}},  \emph {et~al.},\ }\href
  {\doibase 10.1038/nature13792} {\bibfield  {journal} {\bibinfo  {journal}
  {Nature}\ }\textbf {\bibinfo {volume} {514}},\ \bibinfo {pages} {470}
  (\bibinfo {year} {2014})}\BibitemShut {NoStop}%
\bibitem [{\citenamefont {Duerloo}\ \emph {et~al.}(2012)\citenamefont
  {Duerloo}, \citenamefont {Ong},\ and\ \citenamefont
  {Reed}}]{duerloo2012intrinsic}%
  \BibitemOpen
  \bibfield  {author} {\bibinfo {author} {\bibfnamefont {K.-A.~N.}\
  \bibnamefont {Duerloo}}, \bibinfo {author} {\bibfnamefont {M.~T.}\
  \bibnamefont {Ong}}, \ and\ \bibinfo {author} {\bibfnamefont {E.~J.}\
  \bibnamefont {Reed}},\ }\href@noop {} {\bibfield  {journal} {\bibinfo
  {journal} {J. Phys. Chem. Lett.}\ }\textbf {\bibinfo {volume} {3}},\ \bibinfo
  {pages} {2871} (\bibinfo {year} {2012})}\BibitemShut {NoStop}%
\bibitem [{\citenamefont {Hinchet}\ \emph {et~al.}(2018)\citenamefont
  {Hinchet}, \citenamefont {Khan}, \citenamefont {Falconi},\ and\ \citenamefont
  {Kim}}]{hinchet2018piezoelectric}%
  \BibitemOpen
  \bibfield  {author} {\bibinfo {author} {\bibfnamefont {R.}~\bibnamefont
  {Hinchet}}, \bibinfo {author} {\bibfnamefont {U.}~\bibnamefont {Khan}},
  \bibinfo {author} {\bibfnamefont {C.}~\bibnamefont {Falconi}}, \ and\
  \bibinfo {author} {\bibfnamefont {S.-W.}\ \bibnamefont {Kim}},\ }\href@noop
  {} {\bibfield  {journal} {\bibinfo  {journal} {Materials Today}\ }\textbf
  {\bibinfo {volume} {21}},\ \bibinfo {pages} {611} (\bibinfo {year}
  {2018})}\BibitemShut {NoStop}%
\bibitem [{\citenamefont {Majdoub}\ \emph {et~al.}(2008)\citenamefont
  {Majdoub}, \citenamefont {Sharma},\ and\ \citenamefont
  {Cagin}}]{majdoub2008enhanced}%
  \BibitemOpen
  \bibfield  {author} {\bibinfo {author} {\bibfnamefont {M.}~\bibnamefont
  {Majdoub}}, \bibinfo {author} {\bibfnamefont {P.}~\bibnamefont {Sharma}}, \
  and\ \bibinfo {author} {\bibfnamefont {T.}~\bibnamefont {Cagin}},\ }\href
  {\doibase 10.1103/PhysRevB.77.125424} {\bibfield  {journal} {\bibinfo
  {journal} {Phys. Rev. B}\ }\textbf {\bibinfo {volume} {77}},\ \bibinfo
  {pages} {125424} (\bibinfo {year} {2008})}\BibitemShut {NoStop}%
\bibitem [{\citenamefont {He}\ \emph {et~al.}(2018)\citenamefont {He},
  \citenamefont {Javvaji},\ and\ \citenamefont {Zhuang}}]{He2018}%
  \BibitemOpen
  \bibfield  {author} {\bibinfo {author} {\bibfnamefont {B.}~\bibnamefont
  {He}}, \bibinfo {author} {\bibfnamefont {B.}~\bibnamefont {Javvaji}}, \ and\
  \bibinfo {author} {\bibfnamefont {X.}~\bibnamefont {Zhuang}},\ }\href@noop {}
  {\bibfield  {journal} {\bibinfo  {journal} {Physica B: Condensed Matter}\
  }\textbf {\bibinfo {volume} {545}},\ \bibinfo {pages} {527 } (\bibinfo {year}
  {2018})}\BibitemShut {NoStop}%
\bibitem [{\citenamefont {Kvashnin}\ \emph {et~al.}(2015)\citenamefont
  {Kvashnin}, \citenamefont {Sorokin},\ and\ \citenamefont
  {Yakobson}}]{kvashnin2015flexoelectricity}%
  \BibitemOpen
  \bibfield  {author} {\bibinfo {author} {\bibfnamefont {A.~G.}\ \bibnamefont
  {Kvashnin}}, \bibinfo {author} {\bibfnamefont {P.~B.}\ \bibnamefont
  {Sorokin}}, \ and\ \bibinfo {author} {\bibfnamefont {B.~I.}\ \bibnamefont
  {Yakobson}},\ }\href@noop {} {\bibfield  {journal} {\bibinfo  {journal} {J.
  Phys. Chem. Lett.}\ }\textbf {\bibinfo {volume} {6}},\ \bibinfo {pages}
  {2740} (\bibinfo {year} {2015})}\BibitemShut {NoStop}%
\bibitem [{\citenamefont {Olson}\ and\ \citenamefont
  {Sundberg}(1978)}]{olson1978atom}%
  \BibitemOpen
  \bibfield  {author} {\bibinfo {author} {\bibfnamefont {M.~L.}\ \bibnamefont
  {Olson}}\ and\ \bibinfo {author} {\bibfnamefont {K.~R.}\ \bibnamefont
  {Sundberg}},\ }\href {\doibase 10.1063/1.436570} {\bibfield  {journal}
  {\bibinfo  {journal} {J. Chem. Phys.}\ }\textbf {\bibinfo {volume} {69}},\
  \bibinfo {pages} {5400} (\bibinfo {year} {1978})}\BibitemShut {NoStop}%
\bibitem [{\citenamefont {Mayer}(2007)}]{Mayer2007}%
  \BibitemOpen
  \bibfield  {author} {\bibinfo {author} {\bibfnamefont {A.}~\bibnamefont
  {Mayer}},\ }\href {\doibase 10.1103/PhysRevB.75.045407} {\bibfield  {journal}
  {\bibinfo  {journal} {Phys. Rev. B}\ }\textbf {\bibinfo {volume} {75}},\
  \bibinfo {pages} {045407} (\bibinfo {year} {2007})}\BibitemShut {NoStop}%
\bibitem [{\citenamefont {Mayer}(2005)}]{Mayer2005}%
  \BibitemOpen
  \bibfield  {author} {\bibinfo {author} {\bibfnamefont {A.}~\bibnamefont
  {Mayer}},\ }\href {\doibase 10.1103/PhysRevB.71.235333} {\bibfield  {journal}
  {\bibinfo  {journal} {Phys. Rev. B}\ }\textbf {\bibinfo {volume} {71}},\
  \bibinfo {pages} {235333} (\bibinfo {year} {2005})}\BibitemShut {NoStop}%
\bibitem [{\citenamefont {Plimpton}(1995)}]{Plimpton1995}%
  \BibitemOpen
  \bibfield  {author} {\bibinfo {author} {\bibfnamefont {S.}~\bibnamefont
  {Plimpton}},\ }\href {\doibase 10.1006/jcph.1995.1039} {\bibfield  {journal}
  {\bibinfo  {journal} {J. Comp. Phys.}\ }\textbf {\bibinfo {volume} {117}},\
  \bibinfo {pages} {1} (\bibinfo {year} {1995})}\BibitemShut {NoStop}%
\bibitem [{siz()}]{sizeinfo}%
  \BibitemOpen
  \href@noop {} {\emph {\bibinfo {title} {We have not considered the boundary
  atom polarization contribution to the total polarization. Therefore, the
  influence on the total polarization is very weak. It is also noted that the
  flexo coefficients converge to reported values when considered larger size
  systems. However, due to the computational resource limitation, we restricted
  the system size to 80 \AA.}}}\BibitemShut {Stop}%
\bibitem [{\citenamefont {Enyashin}\ and\ \citenamefont
  {Ivanovskii}(2011)}]{enyashin2011graphene}%
  \BibitemOpen
  \bibfield  {author} {\bibinfo {author} {\bibfnamefont {A.~N.}\ \bibnamefont
  {Enyashin}}\ and\ \bibinfo {author} {\bibfnamefont {A.~L.}\ \bibnamefont
  {Ivanovskii}},\ }\href {\doibase 10.1002/pssb.201046583} {\bibfield
  {journal} {\bibinfo  {journal} {Phys. Stat. Solidi (b)}\ }\textbf {\bibinfo
  {volume} {248}},\ \bibinfo {pages} {1879} (\bibinfo {year}
  {2011})}\BibitemShut {NoStop}%
\bibitem [{\citenamefont {Javvaji}\ \emph {et~al.}(2017)\citenamefont
  {Javvaji}, \citenamefont {Shenoy}, \citenamefont {Mahapatra}, \citenamefont
  {Ravikumar}, \citenamefont {Hegde},\ and\ \citenamefont
  {Rizwan}}]{Javvaji2017stable}%
  \BibitemOpen
  \bibfield  {author} {\bibinfo {author} {\bibfnamefont {B.}~\bibnamefont
  {Javvaji}}, \bibinfo {author} {\bibfnamefont {B.~M.}\ \bibnamefont {Shenoy}},
  \bibinfo {author} {\bibfnamefont {D.~R.}\ \bibnamefont {Mahapatra}}, \bibinfo
  {author} {\bibfnamefont {A.}~\bibnamefont {Ravikumar}}, \bibinfo {author}
  {\bibfnamefont {G.~M.}\ \bibnamefont {Hegde}}, \ and\ \bibinfo {author}
  {\bibfnamefont {M.~R.}\ \bibnamefont {Rizwan}},\ }\href@noop {} {\bibfield
  {journal} {\bibinfo  {journal} {Applied Surface Science}\ }\textbf {\bibinfo
  {volume} {414}},\ \bibinfo {pages} {25} (\bibinfo {year} {2017})}\BibitemShut
  {NoStop}%
\bibitem [{\citenamefont {{De Jong}}\ \emph {et~al.}(2015)\citenamefont {{De
  Jong}}, \citenamefont {Chen}, \citenamefont {Geerlings}, \citenamefont
  {Asta},\ and\ \citenamefont {Persson}}]{DeJong2015}%
  \BibitemOpen
  \bibfield  {author} {\bibinfo {author} {\bibfnamefont {M.}~\bibnamefont {{De
  Jong}}}, \bibinfo {author} {\bibfnamefont {W.}~\bibnamefont {Chen}}, \bibinfo
  {author} {\bibfnamefont {H.}~\bibnamefont {Geerlings}}, \bibinfo {author}
  {\bibfnamefont {M.}~\bibnamefont {Asta}}, \ and\ \bibinfo {author}
  {\bibfnamefont {K.~A.}\ \bibnamefont {Persson}},\ }\href {\doibase
  10.1038/sdata.2015.53} {\bibfield  {journal} {\bibinfo  {journal} {Sci.
  Data}\ }\textbf {\bibinfo {volume} {2}},\ \bibinfo {pages} {150053} (\bibinfo
  {year} {2015})}\BibitemShut {NoStop}%
\bibitem [{\citenamefont {Robert}\ and\ \citenamefont
  {Danneau}(2014)}]{Robert2014}%
  \BibitemOpen
  \bibfield  {author} {\bibinfo {author} {\bibfnamefont {P.~T.}\ \bibnamefont
  {Robert}}\ and\ \bibinfo {author} {\bibfnamefont {R.}~\bibnamefont
  {Danneau}},\ }\href {\doibase 10.1088/1367-2630/16/1/013019} {\bibfield
  {journal} {\bibinfo  {journal} {New J. Phys.}\ }\textbf {\bibinfo {volume}
  {16}},\ \bibinfo {pages} {013019} (\bibinfo {year} {2014})}\BibitemShut
  {NoStop}%
\bibitem [{\citenamefont {Surya}\ \emph {et~al.}(2012)\citenamefont {Surya},
  \citenamefont {Iyakutti}, \citenamefont {Mizuseki},\ and\ \citenamefont
  {Kawazoe}}]{Surya2012}%
  \BibitemOpen
  \bibfield  {author} {\bibinfo {author} {\bibfnamefont {V.~J.}\ \bibnamefont
  {Surya}}, \bibinfo {author} {\bibfnamefont {K.}~\bibnamefont {Iyakutti}},
  \bibinfo {author} {\bibfnamefont {H.}~\bibnamefont {Mizuseki}}, \ and\
  \bibinfo {author} {\bibfnamefont {Y.}~\bibnamefont {Kawazoe}},\ }\href
  {\doibase 10.1016/j.commatsci.2012.07.016} {\bibfield  {journal} {\bibinfo
  {journal} {Computational Materials Science}\ }\textbf {\bibinfo {volume}
  {65}},\ \bibinfo {pages} {144} (\bibinfo {year} {2012})}\BibitemShut
  {NoStop}%
\bibitem [{\citenamefont {Dumitric{\u{a}}}\ \emph {et~al.}(2002)\citenamefont
  {Dumitric{\u{a}}}, \citenamefont {Landis},\ and\ \citenamefont
  {Yakobson}}]{dumitricua2002curvature}%
  \BibitemOpen
  \bibfield  {author} {\bibinfo {author} {\bibfnamefont {T.}~\bibnamefont
  {Dumitric{\u{a}}}}, \bibinfo {author} {\bibfnamefont {C.~M.}\ \bibnamefont
  {Landis}}, \ and\ \bibinfo {author} {\bibfnamefont {B.~I.}\ \bibnamefont
  {Yakobson}},\ }\href@noop {} {\bibfield  {journal} {\bibinfo  {journal}
  {Chemical physics letters}\ }\textbf {\bibinfo {volume} {360}},\ \bibinfo
  {pages} {182} (\bibinfo {year} {2002})}\BibitemShut {NoStop}%
\bibitem [{\citenamefont {Nikiforov}\ \emph {et~al.}(2014)\citenamefont
  {Nikiforov}, \citenamefont {Dontsova}, \citenamefont {James},\ and\
  \citenamefont {Dumitric\u{a}}}]{Nikiforov2014}%
  \BibitemOpen
  \bibfield  {author} {\bibinfo {author} {\bibfnamefont {I.}~\bibnamefont
  {Nikiforov}}, \bibinfo {author} {\bibfnamefont {E.}~\bibnamefont {Dontsova}},
  \bibinfo {author} {\bibfnamefont {R.~D.}\ \bibnamefont {James}}, \ and\
  \bibinfo {author} {\bibfnamefont {T.}~\bibnamefont {Dumitric\u{a}}},\ }\href
  {\doibase 10.1103/PhysRevB.89.155437} {\bibfield  {journal} {\bibinfo
  {journal} {Phys. Rev. B}\ }\textbf {\bibinfo {volume} {89}},\ \bibinfo
  {pages} {155437} (\bibinfo {year} {2014})}\BibitemShut {NoStop}%
\bibitem [{\citenamefont {Gleiter}(1987)}]{Gleiter1987}%
  \BibitemOpen
  \bibfield  {author} {\bibinfo {author} {\bibfnamefont {R.}~\bibnamefont
  {Gleiter}},\ }\href {\doibase 10.1351/pac198759121585} {\bibfield  {journal}
  {\bibinfo  {journal} {Pure \& Appl. Chem.}\ }\textbf {\bibinfo {volume}
  {59}},\ \bibinfo {pages} {1585} (\bibinfo {year} {1987})}\BibitemShut
  {NoStop}%
\bibitem [{\citenamefont {Hern{\'{a}}ndez}\ \emph {et~al.}(1998)\citenamefont
  {Hern{\'{a}}ndez}, \citenamefont {Goze}, \citenamefont {Bernier},\ and\
  \citenamefont {Rubio}}]{Hernandez1998}%
  \BibitemOpen
  \bibfield  {author} {\bibinfo {author} {\bibfnamefont {E.}~\bibnamefont
  {Hern{\'{a}}ndez}}, \bibinfo {author} {\bibfnamefont {C.}~\bibnamefont
  {Goze}}, \bibinfo {author} {\bibfnamefont {P.}~\bibnamefont {Bernier}}, \
  and\ \bibinfo {author} {\bibfnamefont {A.}~\bibnamefont {Rubio}},\ }\href
  {\doibase 10.1103/PhysRevLett.80.4502} {\bibfield  {journal} {\bibinfo
  {journal} {Phys. Rev. Lett.}\ }\textbf {\bibinfo {volume} {80}},\ \bibinfo
  {pages} {4502} (\bibinfo {year} {1998})}\BibitemShut {NoStop}%
\bibitem [{\citenamefont {Hern{\'{a}}ndez}\ \emph {et~al.}(1999)\citenamefont
  {Hern{\'{a}}ndez}, \citenamefont {Goze}, \citenamefont {Bernier},\ and\
  \citenamefont {Rubio}}]{Hernandez1999}%
  \BibitemOpen
  \bibfield  {author} {\bibinfo {author} {\bibfnamefont {E.}~\bibnamefont
  {Hern{\'{a}}ndez}}, \bibinfo {author} {\bibfnamefont {C.}~\bibnamefont
  {Goze}}, \bibinfo {author} {\bibfnamefont {P.}~\bibnamefont {Bernier}}, \
  and\ \bibinfo {author} {\bibfnamefont {A.}~\bibnamefont {Rubio}},\ }\href
  {\doibase 10.1007/s003390050890} {\bibfield  {journal} {\bibinfo  {journal}
  {Applied Physics A}\ }\textbf {\bibinfo {volume} {68}},\ \bibinfo {pages}
  {287} (\bibinfo {year} {1999})}\BibitemShut {NoStop}%
\bibitem [{\citenamefont {Podsiad{\l}y-Paszkowska}\ and\ \citenamefont
  {Krawiec}(2017)}]{Podsiady-Paszkowska2017}%
  \BibitemOpen
  \bibfield  {author} {\bibinfo {author} {\bibfnamefont {A.}~\bibnamefont
  {Podsiad{\l}y-Paszkowska}}\ and\ \bibinfo {author} {\bibfnamefont
  {M.}~\bibnamefont {Krawiec}},\ }\href {\doibase 10.1039/C7CP02352A}
  {\bibfield  {journal} {\bibinfo  {journal} {Phys. Chem. Chem. Phys.}\
  }\textbf {\bibinfo {volume} {19}},\ \bibinfo {pages} {14269} (\bibinfo {year}
  {2017})}\BibitemShut {NoStop}%
\bibitem [{\citenamefont {Pike}\ \emph {et~al.}(2017)\citenamefont {Pike},
  \citenamefont {{Van Troeye}}, \citenamefont {Dewandre}, \citenamefont
  {Petretto}, \citenamefont {Gonze}, \citenamefont {Rignanese},\ and\
  \citenamefont {Verstraete}}]{Pike2017}%
  \BibitemOpen
  \bibfield  {author} {\bibinfo {author} {\bibfnamefont {N.~A.}\ \bibnamefont
  {Pike}}, \bibinfo {author} {\bibfnamefont {B.}~\bibnamefont {{Van Troeye}}},
  \bibinfo {author} {\bibfnamefont {A.}~\bibnamefont {Dewandre}}, \bibinfo
  {author} {\bibfnamefont {G.}~\bibnamefont {Petretto}}, \bibinfo {author}
  {\bibfnamefont {X.}~\bibnamefont {Gonze}}, \bibinfo {author} {\bibfnamefont
  {G.~M.}\ \bibnamefont {Rignanese}}, \ and\ \bibinfo {author} {\bibfnamefont
  {M.~J.}\ \bibnamefont {Verstraete}},\ }\href {\doibase
  10.1103/PhysRevB.95.201106} {\bibfield  {journal} {\bibinfo  {journal} {Phys.
  Rev. B}\ }\textbf {\bibinfo {volume} {95}},\ \bibinfo {pages} {201106(R)}
  (\bibinfo {year} {2017})}\BibitemShut {NoStop}%
\bibitem [{\citenamefont {Jiang}\ \emph
  {et~al.}(2013{\natexlab{a}})\citenamefont {Jiang}, \citenamefont {Qi},
  \citenamefont {Park},\ and\ \citenamefont {Rabczuk}}]{Jiang2013}%
  \BibitemOpen
  \bibfield  {author} {\bibinfo {author} {\bibfnamefont {J.~W.}\ \bibnamefont
  {Jiang}}, \bibinfo {author} {\bibfnamefont {Z.}~\bibnamefont {Qi}}, \bibinfo
  {author} {\bibfnamefont {H.~S.}\ \bibnamefont {Park}}, \ and\ \bibinfo
  {author} {\bibfnamefont {T.}~\bibnamefont {Rabczuk}},\ }\href@noop {}
  {\bibfield  {journal} {\bibinfo  {journal} {Nanotechnology}\ }\textbf
  {\bibinfo {volume} {24}},\ \bibinfo {pages} {435705} (\bibinfo {year}
  {2013}{\natexlab{a}})}\BibitemShut {NoStop}%
\bibitem [{\citenamefont {Silberstein}(1917)}]{silberstein1917molecular}%
  \BibitemOpen
  \bibfield  {author} {\bibinfo {author} {\bibfnamefont {L.}~\bibnamefont
  {Silberstein}},\ }\href {\doibase 10.1080/14786440108635618} {\bibfield
  {journal} {\bibinfo  {journal} {The London, Edinburgh, and Dublin
  Philosophical Magazine and Journal of Science}\ }\textbf {\bibinfo {volume}
  {33}},\ \bibinfo {pages} {521} (\bibinfo {year} {1917})}\BibitemShut
  {NoStop}%
\bibitem [{\citenamefont {Thole}(1981)}]{thole1981molecular}%
  \BibitemOpen
  \bibfield  {author} {\bibinfo {author} {\bibfnamefont {B.~T.}\ \bibnamefont
  {Thole}},\ }\href {\doibase 10.1016/0301-0104(81)85176-2} {\bibfield
  {journal} {\bibinfo  {journal} {Chem. Phys.}\ }\textbf {\bibinfo {volume}
  {59}},\ \bibinfo {pages} {341} (\bibinfo {year} {1981})}\BibitemShut
  {NoStop}%
\bibitem [{\citenamefont {Frisch}\ \emph {et~al.}(2016)\citenamefont {Frisch},
  \citenamefont {Trucks}, \citenamefont {Schlegel}, \citenamefont {Scuseria},
  \citenamefont {Robb}, \citenamefont {Cheeseman}, \citenamefont {Scalmani},
  \citenamefont {Barone}, \citenamefont {Petersson}, \citenamefont {Nakatsuji},
  \citenamefont {Li}, \citenamefont {Caricato}, \citenamefont {Marenich},
  \citenamefont {Bloino}, \citenamefont {Janesko}, \citenamefont {Gomperts},
  \citenamefont {Mennucci}, \citenamefont {Hratchian}, \citenamefont {Ortiz},
  \citenamefont {Izmaylov}, \citenamefont {Sonnenberg}, \citenamefont
  {Williams-Young}, \citenamefont {Ding}, \citenamefont {Lipparini},
  \citenamefont {Egidi}, \citenamefont {Goings}, \citenamefont {Peng},
  \citenamefont {Petrone}, \citenamefont {Henderson}, \citenamefont
  {Ranasinghe}, \citenamefont {Zakrzewski}, \citenamefont {Gao}, \citenamefont
  {Rega}, \citenamefont {Zheng}, \citenamefont {Liang}, \citenamefont {Hada},
  \citenamefont {Ehara}, \citenamefont {Toyota}, \citenamefont {Fukuda},
  \citenamefont {Hasegawa}, \citenamefont {Ishida}, \citenamefont {Nakajima},
  \citenamefont {Honda}, \citenamefont {Kitao}, \citenamefont {Nakai},
  \citenamefont {Vreven}, \citenamefont {Throssell}, \citenamefont
  {Montgomery}, \citenamefont {Peralta}, \citenamefont {Ogliaro}, \citenamefont
  {Bearpark}, \citenamefont {Heyd}, \citenamefont {Brothers}, \citenamefont
  {Kudin}, \citenamefont {Staroverov}, \citenamefont {Keith}, \citenamefont
  {Kobayashi}, \citenamefont {Normand}, \citenamefont {Raghavachari},
  \citenamefont {Rendell}, \citenamefont {Burant}, \citenamefont {Iyengar},
  \citenamefont {Tomasi}, \citenamefont {Cossi}, \citenamefont {Millam},
  \citenamefont {Klene}, \citenamefont {Adamo}, \citenamefont {Cammi},
  \citenamefont {Ochterski}, \citenamefont {Martin}, \citenamefont {Morokuma},
  \citenamefont {Farkas}, \citenamefont {Foresman},\ and\ \citenamefont
  {Fox}}]{frisch2016gaussian}%
  \BibitemOpen
  \bibfield  {author} {\bibinfo {author} {\bibfnamefont {M.~J.}\ \bibnamefont
  {Frisch}}, \bibinfo {author} {\bibfnamefont {G.~W.}\ \bibnamefont {Trucks}},
  \bibinfo {author} {\bibfnamefont {H.~B.}\ \bibnamefont {Schlegel}}, \bibinfo
  {author} {\bibfnamefont {G.~E.}\ \bibnamefont {Scuseria}}, \bibinfo {author}
  {\bibfnamefont {M.~A.}\ \bibnamefont {Robb}}, \bibinfo {author}
  {\bibfnamefont {J.~R.}\ \bibnamefont {Cheeseman}}, \bibinfo {author}
  {\bibfnamefont {G.}~\bibnamefont {Scalmani}}, \bibinfo {author}
  {\bibfnamefont {V.}~\bibnamefont {Barone}}, \bibinfo {author} {\bibfnamefont
  {G.~A.}\ \bibnamefont {Petersson}}, \bibinfo {author} {\bibfnamefont
  {H.}~\bibnamefont {Nakatsuji}}, \bibinfo {author} {\bibfnamefont
  {X.}~\bibnamefont {Li}}, \bibinfo {author} {\bibfnamefont {M.}~\bibnamefont
  {Caricato}}, \bibinfo {author} {\bibfnamefont {A.~V.}\ \bibnamefont
  {Marenich}}, \bibinfo {author} {\bibfnamefont {J.}~\bibnamefont {Bloino}},
  \bibinfo {author} {\bibfnamefont {B.~G.}\ \bibnamefont {Janesko}}, \bibinfo
  {author} {\bibfnamefont {R.}~\bibnamefont {Gomperts}}, \bibinfo {author}
  {\bibfnamefont {B.}~\bibnamefont {Mennucci}}, \bibinfo {author}
  {\bibfnamefont {H.~P.}\ \bibnamefont {Hratchian}}, \bibinfo {author}
  {\bibfnamefont {J.~V.}\ \bibnamefont {Ortiz}}, \bibinfo {author}
  {\bibfnamefont {A.~F.}\ \bibnamefont {Izmaylov}}, \bibinfo {author}
  {\bibfnamefont {J.~L.}\ \bibnamefont {Sonnenberg}}, \bibinfo {author}
  {\bibfnamefont {D.}~\bibnamefont {Williams-Young}}, \bibinfo {author}
  {\bibfnamefont {F.}~\bibnamefont {Ding}}, \bibinfo {author} {\bibfnamefont
  {F.}~\bibnamefont {Lipparini}}, \bibinfo {author} {\bibfnamefont
  {F.}~\bibnamefont {Egidi}}, \bibinfo {author} {\bibfnamefont
  {J.}~\bibnamefont {Goings}}, \bibinfo {author} {\bibfnamefont
  {B.}~\bibnamefont {Peng}}, \bibinfo {author} {\bibfnamefont {A.}~\bibnamefont
  {Petrone}}, \bibinfo {author} {\bibfnamefont {T.}~\bibnamefont {Henderson}},
  \bibinfo {author} {\bibfnamefont {D.}~\bibnamefont {Ranasinghe}}, \bibinfo
  {author} {\bibfnamefont {V.~G.}\ \bibnamefont {Zakrzewski}}, \bibinfo
  {author} {\bibfnamefont {J.}~\bibnamefont {Gao}}, \bibinfo {author}
  {\bibfnamefont {N.}~\bibnamefont {Rega}}, \bibinfo {author} {\bibfnamefont
  {G.}~\bibnamefont {Zheng}}, \bibinfo {author} {\bibfnamefont
  {W.}~\bibnamefont {Liang}}, \bibinfo {author} {\bibfnamefont
  {M.}~\bibnamefont {Hada}}, \bibinfo {author} {\bibfnamefont {M.}~\bibnamefont
  {Ehara}}, \bibinfo {author} {\bibfnamefont {K.}~\bibnamefont {Toyota}},
  \bibinfo {author} {\bibfnamefont {R.}~\bibnamefont {Fukuda}}, \bibinfo
  {author} {\bibfnamefont {J.}~\bibnamefont {Hasegawa}}, \bibinfo {author}
  {\bibfnamefont {M.}~\bibnamefont {Ishida}}, \bibinfo {author} {\bibfnamefont
  {T.}~\bibnamefont {Nakajima}}, \bibinfo {author} {\bibfnamefont
  {Y.}~\bibnamefont {Honda}}, \bibinfo {author} {\bibfnamefont
  {O.}~\bibnamefont {Kitao}}, \bibinfo {author} {\bibfnamefont
  {H.}~\bibnamefont {Nakai}}, \bibinfo {author} {\bibfnamefont
  {T.}~\bibnamefont {Vreven}}, \bibinfo {author} {\bibfnamefont
  {K.}~\bibnamefont {Throssell}}, \bibinfo {author} {\bibfnamefont {J.~A.}\
  \bibnamefont {Montgomery}, \bibfnamefont {{Jr.}}}, \bibinfo {author}
  {\bibfnamefont {J.~E.}\ \bibnamefont {Peralta}}, \bibinfo {author}
  {\bibfnamefont {F.}~\bibnamefont {Ogliaro}}, \bibinfo {author} {\bibfnamefont
  {M.~J.}\ \bibnamefont {Bearpark}}, \bibinfo {author} {\bibfnamefont {J.~J.}\
  \bibnamefont {Heyd}}, \bibinfo {author} {\bibfnamefont {E.~N.}\ \bibnamefont
  {Brothers}}, \bibinfo {author} {\bibfnamefont {K.~N.}\ \bibnamefont {Kudin}},
  \bibinfo {author} {\bibfnamefont {V.~N.}\ \bibnamefont {Staroverov}},
  \bibinfo {author} {\bibfnamefont {T.~A.}\ \bibnamefont {Keith}}, \bibinfo
  {author} {\bibfnamefont {R.}~\bibnamefont {Kobayashi}}, \bibinfo {author}
  {\bibfnamefont {J.}~\bibnamefont {Normand}}, \bibinfo {author} {\bibfnamefont
  {K.}~\bibnamefont {Raghavachari}}, \bibinfo {author} {\bibfnamefont {A.~P.}\
  \bibnamefont {Rendell}}, \bibinfo {author} {\bibfnamefont {J.~C.}\
  \bibnamefont {Burant}}, \bibinfo {author} {\bibfnamefont {S.~S.}\
  \bibnamefont {Iyengar}}, \bibinfo {author} {\bibfnamefont {J.}~\bibnamefont
  {Tomasi}}, \bibinfo {author} {\bibfnamefont {M.}~\bibnamefont {Cossi}},
  \bibinfo {author} {\bibfnamefont {J.~M.}\ \bibnamefont {Millam}}, \bibinfo
  {author} {\bibfnamefont {M.}~\bibnamefont {Klene}}, \bibinfo {author}
  {\bibfnamefont {C.}~\bibnamefont {Adamo}}, \bibinfo {author} {\bibfnamefont
  {R.}~\bibnamefont {Cammi}}, \bibinfo {author} {\bibfnamefont {J.~W.}\
  \bibnamefont {Ochterski}}, \bibinfo {author} {\bibfnamefont {R.~L.}\
  \bibnamefont {Martin}}, \bibinfo {author} {\bibfnamefont {K.}~\bibnamefont
  {Morokuma}}, \bibinfo {author} {\bibfnamefont {O.}~\bibnamefont {Farkas}},
  \bibinfo {author} {\bibfnamefont {J.~B.}\ \bibnamefont {Foresman}}, \ and\
  \bibinfo {author} {\bibfnamefont {D.~J.}\ \bibnamefont {Fox}},\ }\href@noop
  {} {\enquote {\bibinfo {title} {Gaussian 16 {R}evision {B}.01},}\ } (\bibinfo
  {year} {2016}),\ \bibinfo {note} {gaussian Inc. Wallingford CT}\BibitemShut
  {NoStop}%
\bibitem [{\citenamefont {Olsen}\ and\ \citenamefont
  {J{\o}rgensen}(1985)}]{Olsen1985}%
  \BibitemOpen
  \bibfield  {author} {\bibinfo {author} {\bibfnamefont {J.}~\bibnamefont
  {Olsen}}\ and\ \bibinfo {author} {\bibfnamefont {P.}~\bibnamefont
  {J{\o}rgensen}},\ }\href {\doibase 10.1063/1.448223} {\bibfield  {journal}
  {\bibinfo  {journal} {J. Chem. Phys.}\ }\textbf {\bibinfo {volume} {82}},\
  \bibinfo {pages} {3235} (\bibinfo {year} {1985})}\BibitemShut {NoStop}%
\bibitem [{\citenamefont {Sekino}\ and\ \citenamefont
  {Bartlett}(1986)}]{Sekino1986}%
  \BibitemOpen
  \bibfield  {author} {\bibinfo {author} {\bibfnamefont {H.}~\bibnamefont
  {Sekino}}\ and\ \bibinfo {author} {\bibfnamefont {R.~J.}\ \bibnamefont
  {Bartlett}},\ }\href {\doibase 10.1063/1.451255} {\bibfield  {journal}
  {\bibinfo  {journal} {J. Chem. Phys.}\ }\textbf {\bibinfo {volume} {85}},\
  \bibinfo {pages} {976} (\bibinfo {year} {1986})}\BibitemShut {NoStop}%
\bibitem [{\citenamefont {Ishigami}\ \emph {et~al.}(2007)\citenamefont
  {Ishigami}, \citenamefont {Chen}, \citenamefont {Cullen}, \citenamefont
  {Fuhrer},\ and\ \citenamefont {Williams}}]{Ishigami2007}%
  \BibitemOpen
  \bibfield  {author} {\bibinfo {author} {\bibfnamefont {M.}~\bibnamefont
  {Ishigami}}, \bibinfo {author} {\bibfnamefont {J.~H.}\ \bibnamefont {Chen}},
  \bibinfo {author} {\bibfnamefont {W.~G.}\ \bibnamefont {Cullen}}, \bibinfo
  {author} {\bibfnamefont {M.~S.}\ \bibnamefont {Fuhrer}}, \ and\ \bibinfo
  {author} {\bibfnamefont {E.~D.}\ \bibnamefont {Williams}},\ }\href {\doibase
  10.1021/nl070613a} {\bibfield  {journal} {\bibinfo  {journal} {Nano. Lett.}\
  }\textbf {\bibinfo {volume} {7}},\ \bibinfo {pages} {1643} (\bibinfo {year}
  {2007})}\BibitemShut {NoStop}%
\bibitem [{\citenamefont {Brenner}\ \emph {et~al.}(2002)\citenamefont
  {Brenner}, \citenamefont {Shenderova}, \citenamefont {Harrison},
  \citenamefont {Stuart}, \citenamefont {Ni},\ and\ \citenamefont
  {Sinnott}}]{brenner2002second}%
  \BibitemOpen
  \bibfield  {author} {\bibinfo {author} {\bibfnamefont {D.~W.}\ \bibnamefont
  {Brenner}}, \bibinfo {author} {\bibfnamefont {O.~A.}\ \bibnamefont
  {Shenderova}}, \bibinfo {author} {\bibfnamefont {J.~A.}\ \bibnamefont
  {Harrison}}, \bibinfo {author} {\bibfnamefont {S.~J.}\ \bibnamefont
  {Stuart}}, \bibinfo {author} {\bibfnamefont {B.}~\bibnamefont {Ni}}, \ and\
  \bibinfo {author} {\bibfnamefont {S.~B.}\ \bibnamefont {Sinnott}},\ }\href
  {\doibase 10.1088/0953-8984/14/4/312} {\bibfield  {journal} {\bibinfo
  {journal} {J .Phys: Condens. Matter}\ }\textbf {\bibinfo {volume} {14}},\
  \bibinfo {pages} {783} (\bibinfo {year} {2002})}\BibitemShut {NoStop}%
\bibitem [{\citenamefont {Verma}\ \emph {et~al.}(2007)\citenamefont {Verma},
  \citenamefont {Jindal},\ and\ \citenamefont {Dharamvir}}]{verma2007elastic}%
  \BibitemOpen
  \bibfield  {author} {\bibinfo {author} {\bibfnamefont {V.}~\bibnamefont
  {Verma}}, \bibinfo {author} {\bibfnamefont {V.}~\bibnamefont {Jindal}}, \
  and\ \bibinfo {author} {\bibfnamefont {K.}~\bibnamefont {Dharamvir}},\ }\href
  {\doibase 10.1088/0957-4484/18/43/435711} {\bibfield  {journal} {\bibinfo
  {journal} {Nanotechnology}\ }\textbf {\bibinfo {volume} {18}},\ \bibinfo
  {pages} {435711} (\bibinfo {year} {2007})}\BibitemShut {NoStop}%
\bibitem [{\citenamefont {Abadi}\ \emph {et~al.}(2018)\citenamefont {Abadi},
  \citenamefont {Shirazi}, \citenamefont {Izadifar}, \citenamefont {Sepahi},\
  and\ \citenamefont {Rabczuk}}]{abadi2018fabrication}%
  \BibitemOpen
  \bibfield  {author} {\bibinfo {author} {\bibfnamefont {R.}~\bibnamefont
  {Abadi}}, \bibinfo {author} {\bibfnamefont {A.~H.~N.}\ \bibnamefont
  {Shirazi}}, \bibinfo {author} {\bibfnamefont {M.}~\bibnamefont {Izadifar}},
  \bibinfo {author} {\bibfnamefont {M.}~\bibnamefont {Sepahi}}, \ and\ \bibinfo
  {author} {\bibfnamefont {T.}~\bibnamefont {Rabczuk}},\ }\href {\doibase
  10.1016/j.commatsci.2017.12.022} {\bibfield  {journal} {\bibinfo  {journal}
  {Computational Materials Science}\ }\textbf {\bibinfo {volume} {145}},\
  \bibinfo {pages} {280} (\bibinfo {year} {2018})}\BibitemShut {NoStop}%
\bibitem [{\citenamefont {Zhao}\ \emph {et~al.}(2016)\citenamefont {Zhao},
  \citenamefont {Xu}, \citenamefont {Wang},\ and\ \citenamefont
  {Lin}}]{zhao2016probing}%
  \BibitemOpen
  \bibfield  {author} {\bibinfo {author} {\bibfnamefont {L.}~\bibnamefont
  {Zhao}}, \bibinfo {author} {\bibfnamefont {S.}~\bibnamefont {Xu}}, \bibinfo
  {author} {\bibfnamefont {M.}~\bibnamefont {Wang}}, \ and\ \bibinfo {author}
  {\bibfnamefont {S.}~\bibnamefont {Lin}},\ }\href {\doibase
  10.1021/acs.jpcc.6b09706} {\bibfield  {journal} {\bibinfo  {journal} {J.
  Phys. Chem. C}\ }\textbf {\bibinfo {volume} {120}},\ \bibinfo {pages} {27675}
  (\bibinfo {year} {2016})}\BibitemShut {NoStop}%
\bibitem [{\citenamefont {Onen}\ \emph {et~al.}(2016)\citenamefont {Onen},
  \citenamefont {Kecik}, \citenamefont {Durgun},\ and\ \citenamefont
  {Ciraci}}]{onen2016gan}%
  \BibitemOpen
  \bibfield  {author} {\bibinfo {author} {\bibfnamefont {A.}~\bibnamefont
  {Onen}}, \bibinfo {author} {\bibfnamefont {D.}~\bibnamefont {Kecik}},
  \bibinfo {author} {\bibfnamefont {E.}~\bibnamefont {Durgun}}, \ and\ \bibinfo
  {author} {\bibfnamefont {S.}~\bibnamefont {Ciraci}},\ }\href {\doibase
  10.1103/PhysRevB.93.085431} {\bibfield  {journal} {\bibinfo  {journal} {Phy.
  Rev. B}\ }\textbf {\bibinfo {volume} {93}},\ \bibinfo {pages} {085431}
  (\bibinfo {year} {2016})}\BibitemShut {NoStop}%
\bibitem [{\citenamefont {Nord}\ \emph {et~al.}(2003)\citenamefont {Nord},
  \citenamefont {Albe}, \citenamefont {Erhart},\ and\ \citenamefont
  {Nordlund}}]{nord2003modelling}%
  \BibitemOpen
  \bibfield  {author} {\bibinfo {author} {\bibfnamefont {J.}~\bibnamefont
  {Nord}}, \bibinfo {author} {\bibfnamefont {K.}~\bibnamefont {Albe}}, \bibinfo
  {author} {\bibfnamefont {P.}~\bibnamefont {Erhart}}, \ and\ \bibinfo {author}
  {\bibfnamefont {K.}~\bibnamefont {Nordlund}},\ }\href {\doibase
  10.1088/0953-8984/15/32/324} {\bibfield  {journal} {\bibinfo  {journal} {J
  Phys: Condens. Matter}\ }\textbf {\bibinfo {volume} {15}},\ \bibinfo {pages}
  {5649} (\bibinfo {year} {2003})}\BibitemShut {NoStop}%
\bibitem [{\citenamefont {Dimoulas}(2015)}]{dimoulas2015silicene}%
  \BibitemOpen
  \bibfield  {author} {\bibinfo {author} {\bibfnamefont {A.}~\bibnamefont
  {Dimoulas}},\ }\href {\doibase 10.1016/j.mee.2014.08.013} {\bibfield
  {journal} {\bibinfo  {journal} {Microelectronic engineering}\ }\textbf
  {\bibinfo {volume} {131}},\ \bibinfo {pages} {68} (\bibinfo {year}
  {2015})}\BibitemShut {NoStop}%
\bibitem [{\citenamefont {Padilha}\ and\ \citenamefont
  {Pontes}(2015)}]{padilha2015free}%
  \BibitemOpen
  \bibfield  {author} {\bibinfo {author} {\bibfnamefont {J.~E.}\ \bibnamefont
  {Padilha}}\ and\ \bibinfo {author} {\bibfnamefont {R.~B.}\ \bibnamefont
  {Pontes}},\ }\href {\doibase 10.1021/jp512489m} {\bibfield  {journal}
  {\bibinfo  {journal} {J. Phys. Chem. C}\ }\textbf {\bibinfo {volume} {119}},\
  \bibinfo {pages} {3818} (\bibinfo {year} {2015})}\BibitemShut {NoStop}%
\bibitem [{\citenamefont {Stillinger}\ and\ \citenamefont
  {Weber}(1985)}]{stillinger1985computer}%
  \BibitemOpen
  \bibfield  {author} {\bibinfo {author} {\bibfnamefont {F.~H.}\ \bibnamefont
  {Stillinger}}\ and\ \bibinfo {author} {\bibfnamefont {T.~A.}\ \bibnamefont
  {Weber}},\ }\href {\doibase 10.1103/PhysRevB.31.5262} {\bibfield  {journal}
  {\bibinfo  {journal} {Phys. Rev. B}\ }\textbf {\bibinfo {volume} {31}},\
  \bibinfo {pages} {5262} (\bibinfo {year} {1985})}\BibitemShut {NoStop}%
\bibitem [{\citenamefont {D{\'a}vila}\ and\ \citenamefont
  {Le~Lay}(2016)}]{Davila2016few}%
  \BibitemOpen
  \bibfield  {author} {\bibinfo {author} {\bibfnamefont {M.~E.}\ \bibnamefont
  {D{\'a}vila}}\ and\ \bibinfo {author} {\bibfnamefont {G.}~\bibnamefont
  {Le~Lay}},\ }\href {\doibase 10.1038/srep20714} {\bibfield  {journal}
  {\bibinfo  {journal} {Sci. Rep.}\ }\textbf {\bibinfo {volume} {6}},\ \bibinfo
  {pages} {20714} (\bibinfo {year} {2016})}\BibitemShut {NoStop}%
\bibitem [{\citenamefont {Mahdizadeh}\ and\ \citenamefont
  {Akhlamadi}(2017)}]{mahdizadeh2017optimized}%
  \BibitemOpen
  \bibfield  {author} {\bibinfo {author} {\bibfnamefont {S.~J.}\ \bibnamefont
  {Mahdizadeh}}\ and\ \bibinfo {author} {\bibfnamefont {G.}~\bibnamefont
  {Akhlamadi}},\ }\href {\doibase 10.1016/j.jmgm.2016.11.009} {\bibfield
  {journal} {\bibinfo  {journal} {Journal of Molecular Graphics and Modelling}\
  }\textbf {\bibinfo {volume} {72}},\ \bibinfo {pages} {1} (\bibinfo {year}
  {2017})}\BibitemShut {NoStop}%
\bibitem [{\citenamefont {Chen}\ \emph {et~al.}(2016)\citenamefont {Chen},
  \citenamefont {Meng}, \citenamefont {Jiang}, \citenamefont {Liang},
  \citenamefont {Yang}, \citenamefont {Tan}, \citenamefont {Sun}, \citenamefont
  {Zhang},\ and\ \citenamefont {Ren}}]{chen2016electronic}%
  \BibitemOpen
  \bibfield  {author} {\bibinfo {author} {\bibfnamefont {X.}~\bibnamefont
  {Chen}}, \bibinfo {author} {\bibfnamefont {R.}~\bibnamefont {Meng}}, \bibinfo
  {author} {\bibfnamefont {J.}~\bibnamefont {Jiang}}, \bibinfo {author}
  {\bibfnamefont {Q.}~\bibnamefont {Liang}}, \bibinfo {author} {\bibfnamefont
  {Q.}~\bibnamefont {Yang}}, \bibinfo {author} {\bibfnamefont {C.}~\bibnamefont
  {Tan}}, \bibinfo {author} {\bibfnamefont {X.}~\bibnamefont {Sun}}, \bibinfo
  {author} {\bibfnamefont {S.}~\bibnamefont {Zhang}}, \ and\ \bibinfo {author}
  {\bibfnamefont {T.}~\bibnamefont {Ren}},\ }\href {\doibase
  10.1039/c6cp02424f} {\bibfield  {journal} {\bibinfo  {journal} {Phys. Chem.
  Chem. Phys.}\ }\textbf {\bibinfo {volume} {18}},\ \bibinfo {pages} {16302}
  (\bibinfo {year} {2016})}\BibitemShut {NoStop}%
\bibitem [{\citenamefont {Saxena}\ \emph {et~al.}(2016)\citenamefont {Saxena},
  \citenamefont {Chaudhary},\ and\ \citenamefont {Shukla}}]{saxena2016stanene}%
  \BibitemOpen
  \bibfield  {author} {\bibinfo {author} {\bibfnamefont {S.}~\bibnamefont
  {Saxena}}, \bibinfo {author} {\bibfnamefont {R.~P.}\ \bibnamefont
  {Chaudhary}}, \ and\ \bibinfo {author} {\bibfnamefont {S.}~\bibnamefont
  {Shukla}},\ }\href {\doibase 10.1038/srep31073} {\bibfield  {journal}
  {\bibinfo  {journal} {Sci. Rep.}\ }\textbf {\bibinfo {volume} {6}},\ \bibinfo
  {pages} {31073} (\bibinfo {year} {2016})}\BibitemShut {NoStop}%
\bibitem [{\citenamefont {Stewart}\ and\ \citenamefont
  {Spearot}(2013)}]{stewart2013atomistic}%
  \BibitemOpen
  \bibfield  {author} {\bibinfo {author} {\bibfnamefont {J.~A.}\ \bibnamefont
  {Stewart}}\ and\ \bibinfo {author} {\bibfnamefont {D.}~\bibnamefont
  {Spearot}},\ }\href {\doibase 10.1088/0965-0393/21/4/045003} {\bibfield
  {journal} {\bibinfo  {journal} {Modell. Simul. Mater. Sci. Eng.}\ }\textbf
  {\bibinfo {volume} {21}},\ \bibinfo {pages} {045003} (\bibinfo {year}
  {2013})}\BibitemShut {NoStop}%
\bibitem [{\citenamefont {Jiang}\ \emph {et~al.}(2016)\citenamefont {Jiang},
  \citenamefont {Jia}, \citenamefont {Zhou}, \citenamefont {Pu}, \citenamefont
  {Zhang},\ and\ \citenamefont {Zhang}}]{jiang2016first}%
  \BibitemOpen
  \bibfield  {author} {\bibinfo {author} {\bibfnamefont {H.-L.}\ \bibnamefont
  {Jiang}}, \bibinfo {author} {\bibfnamefont {S.-H.}\ \bibnamefont {Jia}},
  \bibinfo {author} {\bibfnamefont {D.-W.}\ \bibnamefont {Zhou}}, \bibinfo
  {author} {\bibfnamefont {C.-Y.}\ \bibnamefont {Pu}}, \bibinfo {author}
  {\bibfnamefont {F.-W.}\ \bibnamefont {Zhang}}, \ and\ \bibinfo {author}
  {\bibfnamefont {S.}~\bibnamefont {Zhang}},\ }\href {\doibase
  10.1515/zna-2015-0517} {\bibfield  {journal} {\bibinfo  {journal}
  {Zeitschrift f{\"u}r Naturforschung A}\ }\textbf {\bibinfo {volume} {71}},\
  \bibinfo {pages} {517} (\bibinfo {year} {2016})}\BibitemShut {NoStop}%
\bibitem [{\citenamefont {Jiang}\ \emph
  {et~al.}(2013{\natexlab{b}})\citenamefont {Jiang}, \citenamefont {Park},\
  and\ \citenamefont {Rabczuk}}]{jiang2013molecular}%
  \BibitemOpen
  \bibfield  {author} {\bibinfo {author} {\bibfnamefont {J.-W.}\ \bibnamefont
  {Jiang}}, \bibinfo {author} {\bibfnamefont {H.~S.}\ \bibnamefont {Park}}, \
  and\ \bibinfo {author} {\bibfnamefont {T.}~\bibnamefont {Rabczuk}},\ }\href
  {\doibase 10.1063/1.4818414} {\bibfield  {journal} {\bibinfo  {journal} {J.
  Appl. Phys.}\ }\textbf {\bibinfo {volume} {114}},\ \bibinfo {pages} {064307}
  (\bibinfo {year} {2013}{\natexlab{b}})}\BibitemShut {NoStop}%
\bibitem [{\citenamefont {Wang}\ \emph {et~al.}(2018)\citenamefont {Wang},
  \citenamefont {Bai}, \citenamefont {Yang}, \citenamefont {Fan}, \citenamefont
  {Xie},\ and\ \citenamefont {Li}}]{wang2018electronic}%
  \BibitemOpen
  \bibfield  {author} {\bibinfo {author} {\bibfnamefont {W.}~\bibnamefont
  {Wang}}, \bibinfo {author} {\bibfnamefont {L.}~\bibnamefont {Bai}}, \bibinfo
  {author} {\bibfnamefont {C.}~\bibnamefont {Yang}}, \bibinfo {author}
  {\bibfnamefont {K.}~\bibnamefont {Fan}}, \bibinfo {author} {\bibfnamefont
  {Y.}~\bibnamefont {Xie}}, \ and\ \bibinfo {author} {\bibfnamefont
  {M.}~\bibnamefont {Li}},\ }\href {\doibase 10.3390/ma11020218} {\bibfield
  {journal} {\bibinfo  {journal} {Materials}\ }\textbf {\bibinfo {volume}
  {11}},\ \bibinfo {pages} {218} (\bibinfo {year} {2018})}\BibitemShut
  {NoStop}%
\bibitem [{\citenamefont {Jiang}\ and\ \citenamefont
  {Zhou}(2017)}]{jiang2017parameterization}%
  \BibitemOpen
  \bibfield  {author} {\bibinfo {author} {\bibfnamefont {J.-W.}\ \bibnamefont
  {Jiang}}\ and\ \bibinfo {author} {\bibfnamefont {Y.-P.}\ \bibnamefont
  {Zhou}},\ }in\ \href {\doibase 10.5772/intechopen.71929} {\emph {\bibinfo
  {booktitle} {Handbook of Stillinger-Weber Potential Parameters for
  Two-Dimensional Atomic Crystals}}},\ \bibinfo {editor} {edited by\ \bibinfo
  {editor} {\bibfnamefont {J.-W.}\ \bibnamefont {Jiang}}\ and\ \bibinfo
  {editor} {\bibfnamefont {Y.-P.}\ \bibnamefont {Zhou}}}\ (\bibinfo
  {publisher} {IntechOpen},\ \bibinfo {address} {Rijeka},\ \bibinfo {year}
  {2017})\ Chap.~\bibinfo {chapter} {1}\BibitemShut {NoStop}%
\bibitem [{\citenamefont {Droth}\ \emph {et~al.}(2016)\citenamefont {Droth},
  \citenamefont {Burkard},\ and\ \citenamefont {Pereira}}]{Droth2016}%
  \BibitemOpen
  \bibfield  {author} {\bibinfo {author} {\bibfnamefont {M.}~\bibnamefont
  {Droth}}, \bibinfo {author} {\bibfnamefont {G.}~\bibnamefont {Burkard}}, \
  and\ \bibinfo {author} {\bibfnamefont {V.~M.}\ \bibnamefont {Pereira}},\
  }\href {\doibase 10.1103/PhysRevB.94.075404} {\bibfield  {journal} {\bibinfo
  {journal} {Phys. Rev. B}\ }\textbf {\bibinfo {volume} {94}},\ \bibinfo
  {pages} {075404} (\bibinfo {year} {2016})}\BibitemShut {NoStop}%
\bibitem [{\citenamefont {Zhu}\ \emph {et~al.}(2015)\citenamefont {Zhu},
  \citenamefont {Wang}, \citenamefont {Xiao}, \citenamefont {Liu},
  \citenamefont {Xiong}, \citenamefont {Wong}, \citenamefont {Ye},
  \citenamefont {Ye}, \citenamefont {Yin},\ and\ \citenamefont
  {Zhang}}]{Zhu2015}%
  \BibitemOpen
  \bibfield  {author} {\bibinfo {author} {\bibfnamefont {H.}~\bibnamefont
  {Zhu}}, \bibinfo {author} {\bibfnamefont {Y.}~\bibnamefont {Wang}}, \bibinfo
  {author} {\bibfnamefont {J.}~\bibnamefont {Xiao}}, \bibinfo {author}
  {\bibfnamefont {M.}~\bibnamefont {Liu}}, \bibinfo {author} {\bibfnamefont
  {S.}~\bibnamefont {Xiong}}, \bibinfo {author} {\bibfnamefont {Z.~J.}\
  \bibnamefont {Wong}}, \bibinfo {author} {\bibfnamefont {Z.}~\bibnamefont
  {Ye}}, \bibinfo {author} {\bibfnamefont {Y.}~\bibnamefont {Ye}}, \bibinfo
  {author} {\bibfnamefont {X.}~\bibnamefont {Yin}}, \ and\ \bibinfo {author}
  {\bibfnamefont {X.}~\bibnamefont {Zhang}},\ }\href {\doibase
  10.1038/NNANO.2014.309} {\bibfield  {journal} {\bibinfo  {journal} {Nat.
  Nanotechnol.}\ }\textbf {\bibinfo {volume} {10}},\ \bibinfo {pages} {151}
  (\bibinfo {year} {2015})}\BibitemShut {NoStop}%
\bibitem [{\citenamefont {Mele}\ and\ \citenamefont
  {Kr{\'{a}}l}(2002)}]{Mele2002}%
  \BibitemOpen
  \bibfield  {author} {\bibinfo {author} {\bibfnamefont {E.~J.}\ \bibnamefont
  {Mele}}\ and\ \bibinfo {author} {\bibfnamefont {P.}~\bibnamefont
  {Kr{\'{a}}l}},\ }\href {\doibase 10.1103/PhysRevLett.88.056803} {\bibfield
  {journal} {\bibinfo  {journal} {Phys. Rev. Lett.}\ }\textbf {\bibinfo
  {volume} {88}},\ \bibinfo {pages} {568031} (\bibinfo {year}
  {2002})}\BibitemShut {NoStop}%
\bibitem [{\citenamefont {Sai}\ and\ \citenamefont {Mele}(2003)}]{Sai2003}%
  \BibitemOpen
  \bibfield  {author} {\bibinfo {author} {\bibfnamefont {N.}~\bibnamefont
  {Sai}}\ and\ \bibinfo {author} {\bibfnamefont {E.~J.}\ \bibnamefont {Mele}},\
  }\href {\doibase 10.1103/PhysRevB.68.241405} {\bibfield  {journal} {\bibinfo
  {journal} {Phys. Rev. B}\ }\textbf {\bibinfo {volume} {68}},\ \bibinfo
  {pages} {241405} (\bibinfo {year} {2003})}\BibitemShut {NoStop}%
\bibitem [{\citenamefont {Nakhmanson}\ \emph {et~al.}(2003)\citenamefont
  {Nakhmanson}, \citenamefont {Calzolari}, \citenamefont {Meunier},
  \citenamefont {Bernholc},\ and\ \citenamefont {{Buongiorno
  Nardelli}}}]{Nakhmanson}%
  \BibitemOpen
  \bibfield  {author} {\bibinfo {author} {\bibfnamefont {S.~M.}\ \bibnamefont
  {Nakhmanson}}, \bibinfo {author} {\bibfnamefont {A.}~\bibnamefont
  {Calzolari}}, \bibinfo {author} {\bibfnamefont {V.}~\bibnamefont {Meunier}},
  \bibinfo {author} {\bibfnamefont {J.}~\bibnamefont {Bernholc}}, \ and\
  \bibinfo {author} {\bibfnamefont {M.}~\bibnamefont {{Buongiorno Nardelli}}},\
  }\href {\doibase 10.1103/PhysRevB.67.235406} {\bibfield  {journal} {\bibinfo
  {journal} {Phys. Rev. B}\ }\textbf {\bibinfo {volume} {67}},\ \bibinfo
  {pages} {235406} (\bibinfo {year} {2003})}\BibitemShut {NoStop}%
\bibitem [{\citenamefont {Shu}\ \emph {et~al.}(2011)\citenamefont {Shu},
  \citenamefont {Wei}, \citenamefont {Pang}, \citenamefont {Yao},\ and\
  \citenamefont {Wang}}]{Shu2011}%
  \BibitemOpen
  \bibfield  {author} {\bibinfo {author} {\bibfnamefont {L.}~\bibnamefont
  {Shu}}, \bibinfo {author} {\bibfnamefont {X.}~\bibnamefont {Wei}}, \bibinfo
  {author} {\bibfnamefont {T.}~\bibnamefont {Pang}}, \bibinfo {author}
  {\bibfnamefont {X.}~\bibnamefont {Yao}}, \ and\ \bibinfo {author}
  {\bibfnamefont {C.}~\bibnamefont {Wang}},\ }\href {\doibase
  10.1063/1.3662196} {\bibfield  {journal} {\bibinfo  {journal} {J. Appl.
  Phys.}\ }\textbf {\bibinfo {volume} {110}},\ \bibinfo {pages} {104106}
  (\bibinfo {year} {2011})}\BibitemShut {NoStop}%
\bibitem [{\citenamefont {Ahmadpoor}\ and\ \citenamefont
  {Sharma}(2015)}]{Ahmadpoor2015}%
  \BibitemOpen
  \bibfield  {author} {\bibinfo {author} {\bibfnamefont {F.}~\bibnamefont
  {Ahmadpoor}}\ and\ \bibinfo {author} {\bibfnamefont {P.}~\bibnamefont
  {Sharma}},\ }\href {\doibase 10.1039/C5NR04722F} {\bibfield  {journal}
  {\bibinfo  {journal} {Nanoscale}\ }\textbf {\bibinfo {volume} {7}},\ \bibinfo
  {pages} {16555} (\bibinfo {year} {2015})}\BibitemShut {NoStop}%
\bibitem [{\citenamefont {Naumov}\ \emph {et~al.}(2009)\citenamefont {Naumov},
  \citenamefont {Bratkovsky},\ and\ \citenamefont {Ranjan}}]{Naumov2009}%
  \BibitemOpen
  \bibfield  {author} {\bibinfo {author} {\bibfnamefont {I.}~\bibnamefont
  {Naumov}}, \bibinfo {author} {\bibfnamefont {A.~M.}\ \bibnamefont
  {Bratkovsky}}, \ and\ \bibinfo {author} {\bibfnamefont {V.}~\bibnamefont
  {Ranjan}},\ }\href {\doibase 10.1103/PhysRevLett.102.217601} {\bibfield
  {journal} {\bibinfo  {journal} {Phys. Rev. Lett.}\ }\textbf {\bibinfo
  {volume} {102}},\ \bibinfo {pages} {217601} (\bibinfo {year}
  {2009})}\BibitemShut {NoStop}%
\bibitem [{\citenamefont {Wirtz}\ \emph {et~al.}(2003)\citenamefont {Wirtz},
  \citenamefont {Rubio}, \citenamefont {de~la Concha},\ and\ \citenamefont
  {Loiseau}}]{Wirtz2003}%
  \BibitemOpen
  \bibfield  {author} {\bibinfo {author} {\bibfnamefont {L.}~\bibnamefont
  {Wirtz}}, \bibinfo {author} {\bibfnamefont {A.}~\bibnamefont {Rubio}},
  \bibinfo {author} {\bibfnamefont {R.~A.}\ \bibnamefont {de~la Concha}}, \
  and\ \bibinfo {author} {\bibfnamefont {A.}~\bibnamefont {Loiseau}},\ }\href
  {\doibase 10.1103/PhysRevB.68.045425} {\bibfield  {journal} {\bibinfo
  {journal} {Phys. Rev. B}\ }\textbf {\bibinfo {volume} {68}},\ \bibinfo
  {pages} {045425} (\bibinfo {year} {2003})}\BibitemShut {NoStop}%
\bibitem [{\citenamefont {Moon}\ and\ \citenamefont {Hwang}(2004)}]{Moon2004}%
  \BibitemOpen
  \bibfield  {author} {\bibinfo {author} {\bibfnamefont {W.~H.}\ \bibnamefont
  {Moon}}\ and\ \bibinfo {author} {\bibfnamefont {H.~J.}\ \bibnamefont
  {Hwang}},\ }\href {\doibase 10.1016/j.physe.2003.11.273} {\bibfield
  {journal} {\bibinfo  {journal} {Physica E}\ }\textbf {\bibinfo {volume}
  {23}},\ \bibinfo {pages} {26} (\bibinfo {year} {2004})}\BibitemShut {NoStop}%
\end{thebibliography}%

\end{document}